\documentstyle[preprint,aps,psfig,epsfig]{revtex}
\tighten

\begin{document} 
\def\eqn#1{Eq.$\,$#1}
\def\mb#1{\setbox0=\hbox{$#1$}\kern-.025em\copy0\kern-\wd0
\kern-0.05em\copy0\kern-\wd0\kern-.025em\raise.0233em\box0}
\draft
\preprint{}

\title{Generalized thermodynamics and Fokker-Planck equations. Applications to stellar dynamics and two-dimensional turbulence}
\author{Pierre-Henri Chavanis}
\address{Laboratoire de Physique Th\'eorique, Universit\'e
Paul Sabatier\\ 118 route de Narbonne, 31062 Toulouse Cedex 4,
France\\ (chavanis{@}irsamc2.ups-tlse.fr)}
\maketitle 

\begin{abstract}

We introduce a new set of generalized Fokker-Planck equations that
conserve energy and mass and increase a generalized entropy functional
until a maximum entropy state is reached. Nonlinear Fokker-Planck
equations associated with Tsallis entropies are a special case of
these equations. Application of these results to stellar dynamics and
vortex dynamics are proposed. Our prime result is a novel relaxation
equation that should offer an easily implementable parametrization of
2D turbulence.  Usual parametrizations (including a single turbulent
viscosity) correspond to the infinite temperature limit of our
model. They forget a fundamental systematic drift that acts against
diffusion as in Brownian theory. Our generalized Fokker-Planck
equations can have applications in other fields of physics such as
chemotaxis for bacterial populations. We propose the idea of a
classification of generalized entropies in ``classes of equivalence''
and provide an aesthetic connexion between topics (vortices, stars,
bacteries,...)  which were previously disconnected.

\end{abstract}


\newpage

\section{Introduction}
\label{sec_introduction}

The statistical mechanics of systems with long-range interactions is
currently a topic of active research \cite{dauxois}. Systems with
long-range interactions are numerous in nature: self-gravitating
systems, two-dimensional vortices, non-neutral plasmas, metallic
clusters, dipoles, fracture etc... These systems exhibit similar
features such as negative specific heats, inequivalence of statistical
ensembles, phase transitions and self-organization.  Since they are
non-extensive and non-additive, the construction of an appropriate
thermodynamics is a challenging problem. Among all the previous
examples, self-gravitating systems and 2D vortices play a special role
because they both interact via an unshielded Newtonian potential (in
dimensions $D=3$ or $D=2$) and possess a rather similar mathematical
structure \cite{houches}.

It has been recently argued that the classical Boltzmann entropy may
not be correct for systems with long-range interactions and that
Tsallis entropies, also called $q$-entropies, should be used instead
\cite{tsallis}.  In the context of 2D turbulence, 
Boghosian \cite{boghosian} has interpreted a result of plasma physics
\cite{huang} in terms of Tsallis generalized thermodynamics. In the
astrophysical context, Plastino \& Plastino
\cite{plastino} have noted that the maximization of Tsallis entropies
leads to stellar polytropes, thereby avoiding the {infinite mass
problem} associated with isothermal systems obtained by maximizing the
Boltzmann entropy.  However, the arguments advocated to justify
Tsallis entropies in the context of 2D turbulence and stellar dynamics
are usually unclear and misleading and were criticized in our previous
papers \cite{brands,poly,grand}. In particular, we have argued that
Tsallis entropies are particular $H$-functions (not true entropies)
\cite{tremaine} whose maximization at fixed mass and energy determines
(nonlinearly) dynamically stable stationary solutions of the 2D Euler
or Vlasov-Poisson systems \cite{ipser,ellis}. The $H$-functions can be
useful in describing the metaequilibrium states resulting from
incomplete violent relaxation \cite{lb}.  In this context, the
aforementioned maximization problem is a condition of {dynamical}
stability, not a condition of thermodynamical stability.  Therefore,
Tsallis entropies have not a fundamental justification for stellar
systems and 2D turbulence. They form just a one-parameter family of
$H$-functions that leads to simple models (stellar polytropes and
polytropic vortices). Tsallis distributions can provide, however, a
convenient {\it fit} of the metaequilibrium state in case of
incomplete relaxation. In that context, the parameter $q$, which can
vary in space, measures the importance of mixing \cite{grand}.

On the other hand, it has been shown that Tsallis generalized
thermodynamics could be useful to interpret anomalous diffusion in
complex systems and that the $q$-entropies are connected to nonlinear
Fokker-Planck equations \cite{bukman}. In fact, Tsallis entropies are
just a particular class of a much larger class of functionals that we
shall call {\it generalized entropies}. These functionals are defined
as $S=-\int C(f)d^{3}{\bf r}d^{3}{\bf v}$ where $C(f)$ is a convex
function of the distribution function. Many important properties
obtained with Tsallis $q$-entropies (nonlinear Fokker-Planck
equations, generalized $H$-theorem, Legendre transforms,...)  remain
valid for these more general functionals. Tsallis entropies give a
special importance to power-laws. Power-laws are indeed important in
physics (in relation, among other, with multifractality
\cite{tsallis}) but they are not the most general
distributions. Generalized entropies arise naturally when the
diffusion coefficient is an arbitrary function $D(f)$ of the
distribution function. When the diffusion is counter-balanced by a
friction or a drift, they play the role of Lyapunov functionals and
satisfy a $H$-theorem $\dot S\ge 0$. In this context, Tsallis
entropies correspond to a power-law dependance of the diffusion
coefficient $D(f)\sim f^{q-1}$ and the $q$-parameter in Tsallis formalism
is related to the exponent of anomalous diffusion.

In the first part of the paper, we develop a generalized
thermodynamical {formalism} for a large class of entropy functionals
encompassing Boltzmann, Fermi and Tsallis entropies. This formalism
can have applications in different domains of physics (or biology,
economy, mathematics,...) with different interpretations.  In
Sec. \ref{sec_ge}, we introduce generalized entropies that extend
those introduced by Tsallis and co-workers. In Sec. \ref{sec_ts}, we
establish the conditions of generalized thermodynamical stablity in
microcanonical and canonical ensembles and discuss the possible
inequivalence of statistical ensembles when the caloric curve presents
turning points or bifurcations. In Secs. \ref{sec_kramers} and
\ref{sec_landau}, we introduce generalized kinetic equations
(Fokker-Planck, Landau) that conserve mass and energy and increase a
generalized entropy functional instead of the Boltzmann entropy.  In
Sec. \ref{sec_gsp}, we study the generalized Smoluchowski-Poisson
system and mention possible applications to bacterial populations
(chemotaxis).

In the second part of the paper (Sec. \ref{sec_app}), we discuss the
statistical mechanics of Hamiltonian systems with long-range
interactions focusing on stellar systems and 2D point vortices (or
inviscid continuous vorticity fields). We consider the collisionless
regime when the $N\rightarrow +\infty$ limit is taken before the
$t\rightarrow +\infty$ limit (Vlasov limit). Due to mean-field effects
and long-range interactions, the system undergoes a violent relaxation
(Sec. \ref{sec_vrms}). The resulting {\it metaequilibrium} state is a
stationary solution of the Vlasov (or Euler) equation on the
coarse-grained scale. Its nonlinear dynamical stability can be settled
via a {\it thermodynamical analogy} (Sec. \ref{sec_dsta}). Generalized
entropies (also called H-functions) arise due to the existence of
fine-grained constraints (Casimirs), non-ergodicity and non-ideal
effects (forcing, dissipation,...). We propose a novel relaxation
equation (Sec. \ref{sec_vrds}) that can serve either as a small-scale
parametrization of 2D turbulence (Sec. \ref{sec_relax}) or as a
powerful numerical algorithm to compute arbitrary nonlinearly
dynamically stable stationary solutions of the 2D Euler-Poisson system
(Sec. \ref{sec_na}).  We propose the idea of a classification of
generalized entropies in ``classes of equivalence'' with the heuristic
argument that entropies of the same class should lead to similar
results (Sec. \ref{sec_class}).

\section{Generalized thermodynamics and Fokker-Planck equations}
\label{sec_gt}

\subsection{Generalized entropies}
\label{sec_ge}

Let us consider a system of $N$ particles in interaction and denote by
$f({\bf r},{\bf v},t)$ their distribution function defined such that
$f d^{3}{\bf r}d^{3}{\bf v}$ gives the total mass of particles with
position ${\bf r}$ and velocity ${\bf v}$ at time $t$. Let ${\bf
F}({\bf r},t)=-\nabla\Phi$ be the force (by unit of
mass) experienced by a particle. We assume that the potential $\Phi({\bf
r},t)$ is related to the density $\rho({\bf r},t)=\int f
d^{3}{\bf v}$ by a relation of the form   $\Phi({\bf r})=\int
\rho({\bf r}')u({\bf r}-{\bf r'})d^{3}{\bf r}'$ where $u({\bf r}-{\bf r'})$ is an arbitrary binary potential. For example, if  $u({\bf r}-{\bf r}')=-G/|{\bf r}-{\bf r}'|$, $\Phi$ is solution of the Poisson equation
\begin{equation}
\label{ge1}
\Delta\Phi=4\pi G \rho.
\end{equation}
We assume that the system is isolated so that it conserves mass
\begin{equation}
\label{ge2}
M=\int \rho \ d^{3}{\bf r},
\end{equation}
and energy 
\begin{equation}
\label{ge3}
E=\int{1\over 2}f v^{2} d^{3}{\bf  r}
 d^{3}{\bf  v}+{1\over 2}\int \rho \Phi \ d^{3}{\bf  r}=K+W,
\end{equation} 
where $K$ is the kinetic energy and $W$ the potential energy. The
conservation of angular momentum ${\bf L}=\int f ({\bf r}\times {\bf
v}) d^{3}{\bf r} d^{3}{\bf v}$ and linear impulse ${\bf P}=\int f {\bf
v} d^{3}{\bf r} d^{3}{\bf v}$ can be easily incorporated in the
formalism. The following results remain valid if $\Phi=\Phi_{ext}({\bf
r})$ is a fixed external potential, in which case $W=\int
\rho\Phi_{ext}d^{3}{\bf r}$. 

We introduce a generalized entropy of the form
\begin{equation}
\label{ge4}
S=-\int C(f)d^{3}{\bf r}d^{3}{\bf v},
\end{equation}
where $C(f)$ is a convex function, i.e. $C''(f)> 0$. We are interested by the distribution function $f$ which maximizes the generalized entropy (\ref{ge4}) at fixed mass and energy. Introducing appropriate Lagrange multipliers and writing the variational principle in the form
\begin{equation}
\label{ge5}
\delta S-\beta\delta E-\alpha\delta M=0,
\end{equation}
we find that the critical points of entropy at fixed mass and energy
are given by
\begin{equation}
\label{ge6}
C'(f)=-\beta\epsilon-\alpha,
\end{equation}
where $\epsilon={v^{2}\over 2}+\Phi$ is the energy of a particle by
unit of mass.  The Lagrange multipliers $\beta$ and $\alpha$ are the
generalized inverse temperature and the generalized chemical
potential. Equation (\ref{ge6}) can be written equivalently as
\begin{equation}
\label{ge7}
f=F(\beta\epsilon+\alpha),
\end{equation}
where $F(x)=(C')^{-1}(-x)$. From the identity  
\begin{equation}
\label{ge7b}
f'(\epsilon)=-\beta/C''(f),
\end{equation}
resulting from Eq. (\ref{ge6}), $f(\epsilon)$ is a monotonically decreasing function of energy if $\beta>0$.  The conservation of
angular momentum can be easily included in the variational principle
(\ref{ge5}) by introducing an appropriate Lagrange multiplier ${\bf
\Omega}$. Equation (\ref{ge6}) remains valid provided that $\epsilon$ is
replaced by the Jacobi energy $\epsilon_{J}=\epsilon-{\bf\Omega}\cdot
({\bf r}\times{\bf v})={1\over 2}({\bf v}-{\bf\Omega}\times{\bf
r})^{2}+\Phi_{eff}$ where $\Phi_{eff}=\Phi-{1\over
2}({\bf\Omega}\times {\bf r})^{2}$ is the {\it effective} potential
accounting for inertial forces.

Among all functionals of the form (\ref{ge4}), some have been discussed in detail in the literature. The most famous functional is the Boltzmann entropy
\begin{equation}
\label{bol1}
S_{B}[f]=-\int f\ln f d^{3}{\bf r}d^{3}{\bf v}.
\end{equation}
It leads to the isothermal  (or Boltzmann) distribution 
\begin{equation}
\label{ge8}
f=A e^{-\beta\epsilon}.
\end{equation}
Closely related to the Boltzmann entropy is the Fermi-Dirac entropy 
\begin{equation}
\label{FD1}
S_{F.D.}[f]=-\int \biggl\lbrace {f\over \eta_{0}}\ln {f\over \eta_{0}}+\biggl (1-{f\over \eta_{0}}\biggr )\ln \biggl (1-{f\over \eta_{0}}\biggr ) \biggr\rbrace d^{3}{\bf r} d^{3}{\bf v},
\end{equation}
which leads to the Fermi-Dirac distribution function
\begin{equation}
\label{ge9}
f={\eta_{0}\over 1+\lambda e^{\beta\eta_{0}\epsilon}}.
\end{equation}
The Fermi-Dirac distribution function (\ref{ge9}) satisfies the
constraint $f\le \eta_{0}$ which is related to Pauli's exclusion
principle in quantum mechanics. The isothermal distribution function
(\ref{ge8}) is recovered in the non-degenerate limit $f\ll
\eta_{0}$. We can also consider the case of bosons with the sign $+$
in Eq. (\ref{FD1}).  Recently, there was a considerable interest for
functionals of the form
\begin{equation}
\label{Ts1}
S_{q}[f]=-{1\over q-1}\int (f^{q}-f)  d^{3}{\bf r} d^{3}{\bf v},
\end{equation}
where $q$ is a real number. Such entropies introduced by Tsallis
\cite{tsallis} are called $q$-entropies. They lead to ``polytropic''
distributions of the form
\begin{equation}
\label{ge10}
f=A(\lambda-\epsilon)^{n-{3\over 2}},
\end{equation} 
with $A=\lbrack (q-1)\beta/q\rbrack^{1\over q-1}$ and $\lambda=\lbrack
1-(q-1)\alpha\rbrack /(q-1)\beta$. The index $n$ of the polytrope is
related to the parameter $q$ by the relation $n=3/2+1/(q-1)$.
Isothermal distribution functions are recovered in the limit
$q\rightarrow 1$ (i.e. $n\rightarrow +\infty$).

In any system with $f=f(\epsilon)$, one may define a local energy
dependant excitation temperature by the relation
\begin{equation}
\label{ge11}
{1\over T(\epsilon)}=-{d\ln f\over d\epsilon}.
\end{equation} 
For the isothermal distribution (\ref{ge8}), $T(\epsilon)$ coincides with the
thermodynamic temperature $T=1/\beta$.  For the polytropic distribution
(\ref{ge10}), $T(\epsilon)=(q-1)(\lambda-\epsilon)$. This excitation
temperature has a constant gradient $dT/d\epsilon=q-1$ related to
Tsallis $q$-parameter (or equivalently to the index $n$ of the
polytrope). The other parameter $\lambda$ is related to the value of energy
where the temperature reaches zero.

\subsection{Generalized thermodynamical stability}
\label{sec_ts}

In the preceding section, we have just determined {critical points} of
the generalized entropy (\ref{ge4}) by cancelling its first order
variations with appropriate constraints. We now turn to the
thermodynamical stability of the solutions (in the generalized
sense). We must select {\it maxima} of $S[f]$ at fixed mass and
energy. The condition that $f$ is a {maximum} of $S$ at fixed mass and
energy is equivalent to the condition that $\delta^{2}{J}\equiv
\delta^{2}{S}-\beta\delta^{2}E$ is negative for all perturbations that
conserve mass and energy to first order. This condition can be written
\begin{eqnarray}
\label{ts1}
\delta^{2}J=-\int C''(f){(\delta f)^{2}\over 2}d^{3}{\bf  r} d^{3}{\bf  v}-{1\over 2}\beta\int \delta\rho\delta\Phi d^{3}{\bf  r}\le 0,\nonumber\\
\forall\ \delta f \mid\ \delta E=\delta M=0.\qquad\qquad \qquad
\end{eqnarray}

So far, we have implicitly worked in the {\it microcanonical ensemble} in
which the energy is fixed.  However, it may be of interest to study in
parallel the {\it canonical ensemble} in which the temperature
$T=1/\beta$ is fixed instead of the energy. In that case, the
appropriate thermodynamical potential is the free energy $F=E-TS$ that
we write for convenience in the form of Massieu function
\begin{equation}
\label{ts2}
J=S-\beta E.
\end{equation}
According to Eq. (\ref{ge5}), we have
\begin{equation}
\label{ts3}
\delta J=-E\delta\beta+\alpha\delta M.
\end{equation}
Therefore, the equilibrium state in the canonical ensemble is a
maximum of $J$ at fixed mass and temperature. If we just cancel the
first order variations of $J$, this again yields the relation (\ref{ge6}). The condition of thermodynamical stability in
the canonical ensemble requires that $f$ is a {\it maximum} of $J$ at
fixed mass and temperature. This is equivalent to the condition that
$\delta^{2}{J}$ is negative for
all perturbations that conserve mass. This can be written
\begin{eqnarray}
\label{ts4}
\delta^{2}J=-\int C''(f){(\delta f)^{2}\over 2}d^{3}{\bf  r} d^{3}{\bf  v}-{1\over 2}\beta\int \delta\rho\delta\Phi d^{3}{\bf  r}\le 0,\nonumber\\
\forall\ \delta f \mid\ \delta M=0.\qquad\qquad \qquad
\end{eqnarray}
We note that canonical stability implies microcanonical stability but
the converse is wrong in general. Indeed, if inequality (\ref{ts4}) is
satisfied for all perturbations that conserve mass, it is a fortiori
satisfied for perturbations that conserve mass {\it and} energy. Since
the converse is wrong, this implies that we can ``miss'' some relevant
solutions by working in the canonical ensemble instead of the
microcanonical one.

For self-gravitating systems described by the Boltzmann entropy
(\ref{bol1}), it is
well-known that the statistical ensembles are non-equivalent
\cite{paddy1,houches,grand}. Indeed, an isothermal distribution (\ref{ge8})
can be stable in the microcanonical ensemble (maximum of $S_{B}$ at
fixed $M$ and $E$) but unstable in the canonical ensemble (minimum or
saddle point of $J_{B}$ at fixed $M$ and $T$). In fact, the
inequivalence of statistical ensembles for systems with long-range
interactions is not limited to self-gravitating systems nor to the
Boltzmann entropy (\ref{bol1}). It occurs for many other physical
systems and for various functionals of the form (\ref{ge4}). There
will be inequivalence of statistical ensembles when the caloric curve
$\beta(E)$ presents turning points leading to regions of {\it negative
specific heats}, or said differently, when the entropy $S(E)$ has a
{\it convex dip} \cite{pt,ispolatov}. The stability of the solutions
can be decided by using the turning point criterion of Katz
\cite{katz} which extends the theory of Poincar\'e on linear series of
equilibria. It is found that a change of stability in the series of
equilibria occurs in the microcanonical ensemble when the energy is
extremum and in the canonical ensemble when the temperature is
extremum. Stability is lost or gained depending on whether the series
of equilibria turns clockwise or anti-clockwise at that critical
point. A change of stability along a series of equilibria can also
occur at a branching point \cite{katz,inagaki}, where the solutions bifurcate.
A general classification of phase transitions for systems with long-range interactions has been proposed recently by Bouchet \& Barr\'e \cite{bb}.

\subsection{Generalized Kramers equation}
\label{sec_kramers}

We shall now introduce a generalized Fokker-Planck equation, consistent with 
the thermodynamical framework developed previously, by using a
Maximum Entropy Production Principle \cite{csr}. To apply the MEPP, we  first write the relaxation equation for the distribution function in the form  
\begin{equation}
\label{k1}
{\partial f\over\partial t}+{\bf U}_{6}\cdot \nabla_{6}f=-{\partial {\bf J}_{f}\over \partial {\bf v}},
\end{equation} 
where ${\bf U}_{6}=({\bf v},{\bf F})$ is a generalized velocity field
in the six-dimensional phase space $\lbrace {\bf r},{\bf v}\rbrace$,
$\nabla_{6}=(\partial/\partial {\bf r},\partial/\partial {\bf v})$ is
a generalized gradient and ${\bf J}_{f}$ is the diffusion current to
be determined. The form of Eq. (\ref{k1}) ensures the conservation of
mass provided that ${\bf J}_{f}$ decreases sufficiently rapidly for
large $|{\bf v}|$. From Eqs. (\ref{ge3}), (\ref{ge4}) and (\ref{k1}),
it is easy to put the time variations of energy and entropy in the form
\begin{equation}
\label{k2}
\dot E=\int {\bf J}_{f}\cdot {\bf v} d^{3}{\bf r}d^{3}{\bf v},
\end{equation}
\begin{equation}
\label{k3}
\dot S=-\int C''(f) {\bf J}_{f}\cdot {\partial f\over\partial {\bf v}}d^{3}{\bf r}d^{3}{\bf v},
\end{equation}
where we have used straightforward integrations by parts. Following the MEPP, we shall now determine the optimal current ${\bf J}_{f}$ which maximizes the rate of entropy production (\ref{k3}) while satisfying the conservation of energy $\dot E=0$. For this problem to have a solution, we shall also impose a limitation on the current $|{\bf J}_{f}|$, characterized by a bound $C({\bf r},{\bf v},t)$ which exists but is not known, so that 
\begin{equation}
\label{k4}
{J_{f}^{2}\over 2f}\le C({\bf r},{\bf v},t).
\end{equation}
It can be shown by a convexity argument that reaching the bound (\ref{k4}) is always favorable for increasing $\dot S$, so that this constraint can be replaced by an equality. The variational problem can then be solved by introducing at each time $t$ Lagrange multipliers $\beta$ and $1/D$ for the two constraints. The condition 
\begin{equation}
\label{k5}
\delta\dot S-\beta(t)\delta\dot E-\int {1\over D}\delta \biggl ({J_{f}^{2}\over 2f}\biggr )d^{3}{\bf r}d^{3}{\bf v}=0,
\end{equation}
yields an optimal current of the form
\begin{equation}
\label{k6}
{\bf J}_{f}=-D\biggl\lbrack f C''(f){\partial f\over\partial {\bf v}}+\beta(t)f{\bf v}\biggr \rbrack.
\end{equation}
The time evolution of the Lagrange multiplier $\beta(t)$ is determined by the conservation of energy $\dot E=0$, introducing Eq. (\ref{k6}) in the constraint (\ref{k2}). This yields 
\begin{equation}
\label{k7}
\beta(t)=-{\int D f C''(f){\partial f\over\partial {\bf v}}\cdot {\bf v}d^{3}{\bf r}d^{3}{\bf v}\over \int D f v^{2} d^{3}{\bf r}d^{3}{\bf v}}.  
\end{equation}
Note that the optimal current (\ref{k6}) can be written ${\bf
J}_{f}=Df\partial\alpha/\partial {\bf v}$ where
\begin{equation}
\label{pot}
\alpha({\bf r},{\bf
v},t)\equiv -C'(f)-\beta\epsilon,
\end{equation}
is a generalized potential which is uniform at equilibrium according to 
Eq. (\ref{ge6}). Therefore, the MEPP is just a variational formulation
of the linear thermodynamics of Onsager.

Introducing the optimal current (\ref{k6}) in Eq. (\ref{k1}), we obtain the generalized Fokker-Planck  equation 
\begin{equation}
\label{k8}
{\partial f\over\partial t}+{\bf U}_{6}\cdot \nabla_{6}f={\partial \over \partial {\bf v}}\biggl\lbrace D\biggl\lbrack f C''(f){\partial f\over\partial {\bf v}}+\beta(t)f{\bf v}\biggr \rbrack \biggr\rbrace . 
\end{equation}
Morphologically, Eq.  (\ref{k8}) extends the usual Kramers
equation introduced in the context of colloidal suspensions
\cite{risken} and collisional stellar dynamics \cite{chandra2}. The
first term is a generalized diffusion (depending on the distribution
function) and the second term is a friction. The function $\beta(t)$
can be considered as a time dependant inverse temperature evolving
with time so as to conserve energy (microcanonical formulation). The friction coefficient $\xi=D\beta$ satisfies a generalized Einstein relation. Note that $D$ is not determined by the MEPP since it is related to the unknown bound $C({\bf r},{\bf v},t)$ in Eq. (\ref{k4}). We can use this indetermination to write Eq. (\ref{k8}) in the alternative form
\begin{equation}
\label{k9}
{\partial f\over\partial t}+{\bf U}_{6}\cdot \nabla_{6}f={\partial \over \partial {\bf v}}\biggl\lbrace D'\biggl\lbrack {\partial f\over\partial {\bf v}}+{\beta(t)\over C''(f)}{\bf v}\biggr \rbrack \biggr\rbrace,
\end{equation}
which will have the same general properties as Eq. (\ref{k8}). This equation involves an ordinary diffusion and a nonlinear friction.  Equation (\ref{k9}) can
be deduced from Eq. (\ref{k8}) by the substitution $D'=DfC''(f)$. One
of these two forms will be prefered depending on the situation
contemplated. Note that $D$ or $D'$ can depend on ${\bf r},{\bf v},t$
without altering the general properties of the equations.

It is straightforward to check that Eq. (\ref{k8}) with the constraint
(\ref{k7}) satisfies a $H$-theorem for the generalized entropy
(\ref{ge4}). From Eqs. (\ref{k3}) and (\ref{k6}), we can write
\begin{equation}
\label{k10}
\dot S=-\int {{\bf J}_{f}\over f}\cdot \biggl\lbrack f C''(f){\partial f\over\partial {\bf v}}+\beta(t)f{\bf v}\biggr \rbrack d^{3}{\bf r}d^{3}{\bf v}+\beta(t) \int {\bf J}_{f}\cdot {\bf v} d^{3}{\bf r}d^{3}{\bf v}.
\end{equation}
The last quantity vanishes due to the conservation of energy (\ref{k2}). Therefore,
\begin{equation}
\label{k11}
\dot S=\int {J_{f}^{2}\over Df}d^{3}{\bf r}d^{3}{\bf v},
\end{equation}
which is positive provided that $D>0$. Now, at equilibrium $\dot S=0$, hence ${\bf J}_{f}={\bf 0}$, so that according to Eq. (\ref{k6}),
\begin{equation}
\label{k12}
{\partial C'(f)\over\partial {\bf v}}+\beta {\bf v}={\bf 0}.
\end{equation}
Integrating with respect to ${\bf v}$, we get
\begin{equation}
\label{k13b}
C'(f)=-\beta {v^{2}\over 2}+A({\bf r}).
\end{equation}
The cancellation of the advective term ${\bf U}_{6}\cdot \nabla_{6}$
in Eq. (\ref{k8}) combined with Eq. (\ref{k13b}) implies that
$f=f(\epsilon)$ and $\nabla A=-\beta\nabla\Phi$. Therefore, $A({\bf
r})=-\beta\Phi({\bf r})-\alpha$ and we recover Eq. (\ref{ge6}) with
$\beta=\lim_{t\rightarrow +\infty}\beta(t)$. Therefore, a stationary
solution of Eq. (\ref{k8}) extremizes the entropy at fixed energy and
mass.  In addition, only {\it maxima} of $S$ at fixed $M$ and $E$
are linearly stable with respect to the generalized Fokker-Planck
equation (\ref{k8}). Indeed, considering the linear stability of a stationary solution of Eqs. (\ref{k8}) and (\ref{k7}), we can derive the general relation (see Appendix \ref{sec_lsK})
\begin{eqnarray}
\label{l15}
2\lambda\delta^{2}{J}=\delta^{2}\dot S\ge 0,
\end{eqnarray} 
connecting the growth rate $\lambda$ of the perturbation $\delta f\sim
e^{\lambda t}$ to the second order variations of the free energy
$J=S-\beta E$ and the second order variations of the rate of entropy
production $\delta^{2}\dot S\ge 0$.  Since the product
$\lambda\delta^{2}{J}$ is positive, we conclude that a stationary
solution of the generalized Fokker-Planck equation (\ref{k8}) is
linearly stable ($\lambda<0$) if and only if it is an entropy {\it
maximum } at fixed mass and energy (see Sec. \ref{sec_ts}). This
aesthetic formula shows the equivalence between dynamical and
thermodynamical stability for our generalized Fokker-Planck
equations. Therefore, they only select {\it maxima} of $S$, not minima
or saddle points.

A relaxation equation appropriate to the canonical situation can be obtained by maximizing $\dot J=\dot S-\beta\dot E$ with the constraint (\ref{k4}). The variational principle    
\begin{equation}
\label{k13}
\delta\dot J-\int {1\over D}\delta \biggl ({J_{f}^{2}\over 2f}\biggr )d^{3}{\bf r}d^{3}{\bf v}=0,
\end{equation}
again yields an optimal current of the form (\ref{k6}) but with  constant $\beta$. Since
\begin{equation}
\label{k14}
\dot J=-\int {\bf J}_{f}\cdot \biggl\lbrack  C''(f){\partial f\over\partial {\bf v}}+\beta {\bf v}\biggr \rbrack d^{3}{\bf r}d^{3}{\bf v}=\int {J_{f}^{2}\over Df}d^{3}{\bf r}d^{3}{\bf v}\ge 0,
\end{equation}
according to Eqs. (\ref{k2}), (\ref{k3}) and (\ref{k6}), we find that
the free energy $J$ increases monotonically until an equilibrium state
of the form (\ref{ge6}) is reached. In the canonical ensemble, we can
show that $2\lambda\delta^{2}{J}=\delta^{2}\dot J\ge 0$ and conclude
that a stationary solution of the generalized Fokker-Planck equation
(\ref{k8}) with constant $\beta$ is linearly stable if and only if it
is a {\it maximum} of free energy at fixed mass and temperature
(see Sec. \ref{sec_ts}).

We can also use the MEPP to construct a more
general relaxation equation. Assuming that the diffusion current in
Eq. (\ref{k1}) depends on ${\bf v}$ and ${\bf r}$ and repeating the
same steps as before, we get
\begin{equation}
\label{k8new}
{\partial f\over\partial t}+{\bf U}_{6}\cdot \nabla_{6}f=\nabla_{6}\biggl\lbrace D\biggl\lbrack f C''(f)\nabla_{6} f+\beta(t)f{\bf U}_{6\perp}\biggr \rbrack \biggr\rbrace,
\end{equation}
with
\begin{equation}
\label{k9new}
\beta(t)=-{\int DfC''(f)\nabla_{6}f{\bf U}_{6\perp}d^{3}{\bf r}d^{3}{\bf v}\over \int Df({\bf U}_{6\perp})^{2}d^{3}{\bf r}d^{3}{\bf v}},
\end{equation}
where ${\bf U}_{6\perp}=(-{\bf F},{\bf v})$. The generalized velocity ${\bf U}_{6}$ in phase space is very similar to the velocity field of a two-dimensional incompressible fluid (see Sec. \ref{sec_vrms}). We note in particular that ${\bf U}_{6\perp}=\nabla_{6}\epsilon$, where $\epsilon={v^{2}\over 2}+\Phi$ plays the role of a generalized streamfunction.

To conclude this section, it can be of interest to discuss some
special cases explicitly. For the Boltzmann entropy (\ref{bol1}),
$C''(f)=1/f$ and Eq. (\ref{k8}) has the form of an ordinary Kramers equation
\begin{equation}
\label{k0}
{\partial f\over\partial t}+{\bf U}_{6}\cdot \nabla_{6}f={\partial \over \partial {\bf v}}\biggl\lbrack D\biggl ( {\partial f\over\partial {\bf v}}+\beta f{\bf v}\biggr ) \biggr\rbrack.
\end{equation}
For the Fermi-Dirac entropy (\ref{FD1}), $C''(f)=1/f(\eta_{0}-f)$. In
order to avoid the divergence of the term $f C''(f)$ as $f\rightarrow
\eta_{0}$, it is appropriate to consider the alternative form
(\ref{k9}) of the generalized Kramers equation. This yields
\begin{equation}
\label{k16}
{\partial f\over\partial t}+{\bf U}_{6}\cdot \nabla_{6}f={\partial \over \partial {\bf v}}\biggl\lbrace D'\biggl \lbrack {\partial f\over\partial {\bf v}}+\beta f(\eta_{0}-f){\bf v}\biggr \rbrack \biggr\rbrace,
\end{equation}
which has been initially proposed in \cite{csr}.
Finally, for the Tsallis entropy (\ref{Ts1}), $C''(f)=q
f^{q-2}$ and  Eq. (\ref{k8}) has the form of a
nonlinear Fokker-Planck equation
\begin{equation}
\label{k17}
{\partial f\over\partial t}+{\bf U}_{6}\cdot\nabla_{6}f={\partial \over \partial {\bf v}}\biggl\lbrack D\biggl ( {\partial f^{q}\over\partial {\bf v}}+\beta f {\bf v}\biggr ) \biggr\rbrack . 
\end{equation}
This equation has been studied in detail recently in relation with
Tsallis entropy and anomalous diffusion \cite{bukman}. The underlying
mechanism giving rise to anomalous diffusion may differ depending on
the physical system: L\'evy walkers, porous media, vortex dipoles in 2D
turbulence, etc... In such systems, the diffusion
coefficient $\sim f^{q-1}$ is a power-law of the distribution function
and the phase space has a fractal or multifractal structure (the
exponent $q$ is related to the fractal dimension). In fact, the nice
properties of Eq. (\ref{k17}), in particular the H-theorem, go beyond
the form of entropy considered by Tsallis and remain valid for all
convex function $C(f)$ even if the results are not always explicit. In the context of anomalous diffusion, Eq. (\ref{k8}) can be obtained from a Langevin equation of the form
\begin{equation}
\label{k8a}
{d{\bf v}\over dt}={\bf F}-\xi {\bf v}+\sqrt{2D f\biggl \lbrack {C(f)\over f}\biggr \rbrack'}{\bf R}(t),
\end{equation}
where ${\bf R}(t)$ is a white noise. Since the function in front of
${\bf R}(t)$ depends on $({\bf r},{\bf v})$, the last term in
Eq. (\ref{k8a}) is a multiplicative noise. When $C(f)$ is a power law,
Eq. (\ref{k8a}) reduces to the stochastic equations studied by Borland
\cite{borland}.

\subsection{Generalized Landau equation}
\label{sec_landau}

We shall now introduce another relaxation equation tending to a state
of maximum generalized entropy at fixed mass and energy. This is the
generalized Landau equation
\begin{eqnarray}
\label{m3}
{\partial f\over\partial t}+{\bf U}_{6}\cdot \nabla_{6}f={\partial\over\partial v^{\mu}}\int d^{3}{\bf v}'K^{\mu\nu}f f'\biggl\lbrack C''(f){\partial f\over\partial v^{\nu}}-C''(f'){\partial f'\over\partial v^{'\nu}}\biggr\rbrack,
\end{eqnarray} 
\begin{eqnarray}
\label{m2}
K^{\mu\nu}={A\over u}\biggl (\delta^{\mu\nu}-{u^{\mu}u^{\nu}\over u^{2}}\biggr ),
\end{eqnarray} 
where $f=f({\bf r},{\bf v},t)$, $f'=f({\bf r},{\bf v}',t)$, ${\bf
u}={\bf v}'-{\bf v}$, $A$ is a constant and  $C(f)$ in any convex function. We shall also consider the alternative form
\begin{eqnarray}
\label{m3bis}
{\partial f\over\partial t}+{\bf U}_{6}\cdot \nabla_{6}f={\partial\over\partial v^{\mu}}\int d^{3}{\bf v}'K^{\mu\nu}\biggl\lbrack {1\over C''(f')}{\partial f\over\partial v^{\nu}}-{1\over C''(f)}{\partial f'\over\partial v^{'\nu}}\biggr\rbrack.
\end{eqnarray}
Morphologically, these equations extend the usual Landau equation
introduced in plasma physics and collisional stellar dynamics
\cite{balescu,bt}.  The generalized Landau equation satisfies the
conservation of mass, energy, angular momentum and linear impulse and
increases a generalized entropy (H-theorem). In
addition, formula (\ref{l15}) remains valid so
that a stationary solution of the generalized Landau equation is
linearly stable if and only if it is a maximum of the generalized
entropy (\ref{ge4}). 

Let us consider special cases explicitly. For the Boltzmann entropy,
Eq. (\ref{m3}) reduces to the usual Landau equation
\begin{eqnarray}
\label{m1}
{\partial f\over\partial t}+{\bf U}_{6}\cdot \nabla_{6}f={\partial\over\partial v^{\mu}}\int d^{3}{\bf v}'K^{\mu\nu}\biggl ( f'{\partial f\over\partial v^{\nu}}-f{\partial f'\over\partial v^{'\nu}}\biggr ).
\end{eqnarray} 
For the Fermi-Dirac entropy, Eq. (\ref{m3bis}) takes the form 
\begin{eqnarray}
\label{m4}
{\partial f\over\partial t}+{\bf U}_{6}\cdot \nabla_{6}f={\partial\over\partial v^{\mu}}\int d^{3}{\bf v}'K^{\mu\nu}\biggl \lbrack f'(\eta_{0}-f'){\partial f\over\partial v^{\nu}}-f(\eta_{0}-f){\partial f'\over\partial v^{'\nu}}\biggr \rbrack.
\end{eqnarray}
Finally, for the Tsallis entropy, we get
\begin{eqnarray}
\label{m5}
{\partial f\over\partial t}+{\bf U}_{6}\cdot \nabla_{6}f={\partial\over\partial v^{\mu}}\int d^{3}{\bf v}'K^{\mu\nu}\biggl (f'{\partial f^{q}\over\partial v^{\nu}}-f{\partial f^{'q}\over\partial v^{'\nu}}\biggr ).
\end{eqnarray}
This could be called the $q$-Landau equation. Contrary to the
nonlinear Kramers equation (\ref{k17}), it seems that Eq. (\ref{m5})
has never been introduced previously. We shall study its properties
more specifically in a future work \cite{gLandau}. A connexion between
the (generalized) Landau equation and the (generalized) Kramers
equation can be found in a {\it thermal bath approximation}. In this
context, $f=f({\bf v},t)$ describes a test particle and $f'=f({\bf
v}',t)$ describes the field particles. If we replace $f'$ in
Eq. (\ref{m3}) by its equilibrium value (\ref{ge6}), then
Eq. (\ref{m3}) becomes equivalent to the generalized Kramers equation
(\ref{k8}) and the diffusion coefficient can be calculated \cite{gLandau}.

\section{The generalized Smoluchowski-Poisson system}
\label{sec_gsp}

\subsection{The high friction limit}
\label{sec_xi}

The Kramers-Poisson system (\ref{k0})-(\ref{ge1}) is relatively
complicated because it has to be solved in a six-dimensional phase
space. However, it is well-known in Brownian theory \cite{risken}
that, in the high friction limit $\xi\rightarrow +\infty$ (or
equivalently for large times $t\gg \xi^{-1}$), the velocity
distribution function becomes close to the Maxwellian distribution and
the evolution of the spatial density $\rho({\bf r},t)$ is governed by
the Smoluchowski equation
\begin{eqnarray}
\label{x1}
{\partial\rho\over\partial t}=\nabla \biggl\lbrack {1\over\xi}(T\nabla \rho+\rho\nabla\Phi)\biggr\rbrack.
\end{eqnarray} 
The Smoluchowski-Poisson system has been studied in
\cite{crs,sire}. It models the dynamics of self-gravitating
Brownian particles and the chemotactic aggregation of bacterial
populations.

We now proceed in deriving a generalized Smoluchowski equation by
taking the high friction limit of the generalized Kramers equation. We
shall assume that $\beta$ is constant (canonical situation).
Integrating Eq. (\ref{k8}) over velocity, we get the continuity
equation
\begin{eqnarray}
\label{x2}
{\partial\rho\over\partial t}+\nabla (\rho {\bf u})=0,
\end{eqnarray} 
where ${\bf u}=(1/\rho)\int f{\bf v}d^{3}{\bf v}$ is the local velocity. 
Multiplying Eq. (\ref{k8}) by ${\bf v}$ and integrating over velocity, we get the momentum equation 
\begin{eqnarray}
\label{x3}
{\partial\over\partial t}(\rho u_{i})+{\partial\over\partial x_{j}}(\rho u_{i}u_{j})+{\partial\over\partial x_{j}}P_{ij}+\rho{\partial\Phi\over\partial x_{i}}=-\int D\biggl \lbrack f C''(f){\partial f\over\partial v_{i}}+\beta f v_{i}\biggr \rbrack d^{3}{\bf v},
\end{eqnarray} 
where $P_{ij}=\int f w_{i}w_{j} d^{3}{\bf v}$ is the
stress tensor and ${\bf w}={\bf v}-{\bf u}$ the relative
velocity. Introducing the notation $\phi(f)=\int^{f}xC''(x)dx$, the
first term in the collision term can be rewritten $\partial
\phi(f)/\partial {\bf v}$ and, since it is a gradient of a function,
it vanishes by integration. We are left therefore with
\begin{eqnarray}
\label{x4}
{\partial\over\partial t}(\rho u_{i})+{\partial\over\partial x_{j}}(\rho u_{i}u_{j})+{\partial\over\partial x_{j}}P_{ij}+\rho{\partial\Phi\over\partial x_{i}}=-D\beta\rho  u_{i}.
\end{eqnarray} 
We shall close this hierarchy of equations with the Local Thermodynamical Equilibrium (LTE) condition
\begin{eqnarray}
\label{x5}
C'(f)=-\beta\biggl ({w^{2}\over 2}+\lambda({\bf r},t)\biggr ),
\end{eqnarray} 
where $\lambda({\bf r},t)$ depends only on position and time. This
distribution function maximizes the local density of free energy at
fixed density $\rho$ and velocity ${\bf u}$. The Lagrange multiplier
$\lambda$ is related to the spatial density $\rho({\bf r},t)$ through
the relation
\begin{eqnarray}
\label{x6}
\rho=\int f d^{3}{\bf v}.
\end{eqnarray} 
Furthermore, since the velocity distribution in Eq. (\ref{x5}) is isotropic, we have $P_{ij}=p\delta_{ij}$ where $p({\bf r},t)$ is the local pressure
\begin{eqnarray}
\label{x7}
p={1\over 3}\int f w^{2}d^{3}{\bf v},
\end{eqnarray} 
determined by Eq. (\ref{x5}). From these two relations, we find that
the fluid is {\it barotropic} in the sense that $p({\bf r},t)=p\lbrack
\rho({\bf r},t)\rbrack$ where the function $p[\ ]$ is completely
specified by $C(f)$. We have thus obtained what might be called the
damped Euler-Jeans equations 
\begin{eqnarray}
\label{x8}
{\partial\rho\over\partial t}+\nabla (\rho {\bf u})=0,
\end{eqnarray} 
\begin{eqnarray}
\label{x9}
\rho {d{\bf u}\over dt}=-\nabla p-\rho\nabla\Phi-\xi\rho {\bf u},
\end{eqnarray} 
where $d/dt=\partial/\partial t+{\bf u}\cdot\nabla$ is the material
derivative.  These equations were first proposed in \cite{csr}.
In the high friction limit, these equations can be simplified further since, to first order in $\xi^{-1}$, we have
\begin{eqnarray}
\label{x10}
\rho {\bf u}=-{1\over\xi}(\nabla p+\rho\nabla\Phi),
\end{eqnarray} 
which is obtained from Eq. (\ref{x9}) by neglecting the advective
term. Note that the high friction limit is consistent with the Local
Thermodynamical Equilibrium (\ref{x5}). Indeed, in the high friction limit
$\xi=D\beta\rightarrow +\infty$, the term in bracket in Eq. (\ref{k8})
must vanish so that the distribution function satisfies (\ref{x5}) in
good approximation with ${\bf u}=O(\xi^{-1})$. Inserting the relation
(\ref{x10}) in the continuity equation (\ref{x8}), we get a
generalized form of the Smoluchowski equation
\begin{eqnarray}
\label{x11}
{\partial\rho\over\partial t}=\nabla \biggl\lbrack {1\over\xi}(\nabla p+\rho\nabla\Phi)\biggr\rbrack,
\end{eqnarray} 
initially proposed in \cite{csr}.
The aesthetic form of this equation, in which the pressure $p$
replaces naturally the usual term $\rho T$ in the familiar
Smoluchowski equation, suggests by itself the consistency of the
generalized thermodynamical formalism.   The condition of stationarity in Eq. (\ref{x11})
corresponds to an equilibrium between the pressure force $-\nabla p$
and the mean-field force $-\rho\nabla\Phi$. This is equivalent to
Eq. (\ref{ge6}). Indeed, using Eqs. (\ref{x6}), (\ref{x7}) and
(\ref{ge6}), one has
\begin{eqnarray}
\label{x12}
\rho={1\over 3}\int f{\partial {\bf v}\over\partial {\bf v}}d^{3}{\bf v}=-{1\over 3}\int {\partial f\over\partial {\bf v}}\cdot {\bf v}d^{3}{\bf v}={1\over 3}\beta\int {v^{2}\over C''(f)}d^{3}{\bf v},
\end{eqnarray} 
\begin{eqnarray}
\label{x13}
\nabla p={1\over 3}\int {\partial f\over\partial {\bf r}}v^{2}d^{3}{\bf v}=-{1\over 3}\beta\nabla\Phi\cdot \int {v^{2}\over C''(f)}d^{3}{\bf v},
\end{eqnarray} 
so that we obtain the condition of hydrostatic equilibrium
\begin{eqnarray}
\label{x13b}
\nabla p=-\rho\nabla\Phi.
\end{eqnarray} 
Finally, we can show that the generalized Smoluchowski-Poisson system (\ref{x11}) satisfies a form of Virial theorem
\begin{eqnarray}
\label{x14}
{1\over 2}\xi{dI\over dt}=2K+W-3p_{b}V,
\end{eqnarray} 
where $I=\int \rho r^{2}d^{3}{\bf r}$ is the moment of inertia and
$p_{b}$ the pressure on the box (assumed uniform). The proof is the
same as that given in Appendix D of Ref. \cite{sire}.

To conclude this section, we can consider particular forms of the
generalized Smoluchowski equation. For the Boltzmann entropy (\ref{bol1}),
Eqs. (\ref{x5}), (\ref{x6}) and (\ref{x7}) lead to the isothermal
equation of state $p=\rho T$ and to the usual form (\ref{x1}) of the
Smoluchowski equation.  The case of the
Fermi-Dirac entropy (\ref{FD1}) has been treated in
\cite{csr}. For the Tsallis entropy (\ref{Ts1}), Eqs. (\ref{x5}),
(\ref{x6}) and (\ref{x7}) lead to the polytropic equation of state
$p=K\rho^{\gamma}$ with $\gamma=1+1/n$ and Eq. (\ref{x11}) becomes the
nonlinear Smoluchowski equation
\begin{eqnarray}
\label{x1n}
{\partial\rho\over\partial t}=\nabla \biggl\lbrack {1\over\xi}(K\nabla \rho^{\gamma}+\rho\nabla\Phi)\biggr\rbrack.
\end{eqnarray} 
The nonlinear Smoluchowski-Poisson system has been studied in detail
in Ref. \cite{anomalous}. It describes self-gravitating Langevin
particles experiencing anomalous diffusion. It is likely that
anomalous diffusion occurs also in biological systems so that
nonlinear Smoluchowski equations can find applications in the context
of chemotaxis.

\subsection{The Lyapunov functional}
\label{sec_Lyap}

In Sec. \ref{sec_kramers}, it was indicated that the Lyapunov
functional associated with the generalized Kramers equation (\ref{k8})
with a fixed inverse temperature $\beta$ is the free energy $J=S-\beta
E$ with the generalized entropy (\ref{ge4}). Therefore, the Lyapunov
functional associated with the generalized Smoluchowski equation
(\ref{x11}) will coincide with the simplified form of $J$ obtained by
using the Local Thermodynamical Equilibrium condition (\ref{x5}), with
${\bf u}={\bf 0}$, to express $J[f]$ as a functional of $\rho$. Using
Eq. (\ref{x7}), the energy (\ref{ge3}) can be written
\begin{eqnarray}
\label{L1}
E={3\over 2}\int p d^{3}{\bf r}+{1\over 2}\int\rho\Phi d^{3}{\bf r}.
\end{eqnarray} 
On the other hand, it is shown in \cite{grand} that when $f$ is given by Eq. (\ref{x5}), the entropy (\ref{ge4}) can be rewritten as
\begin{eqnarray}
\label{L2}
S={5\over 2}\beta\int p d^{3}{\bf r}+\beta\int\lambda\rho d^{3}{\bf r}.
\end{eqnarray} 
Therefore, the free energy reads
\begin{eqnarray}
\label{L3}
J=\beta\int p d^{3}{\bf r}+\beta\int\lambda\rho d^{3}{\bf r}-{1\over 2}\beta\int\rho\Phi d^{3}{\bf r}.
\end{eqnarray} 
It is shown furthermore in \cite{grand} that Eq. (\ref{L3}) can be put in the equivalent form
\begin{eqnarray}
\label{L4}
J=-\beta\int \rho\int_{0}^{\rho}{p(\rho')\over \rho'^{2}}d\rho' d^{3}{\bf r}-{1\over 2}\beta\int \rho\Phi d^{3}{\bf r}+{\rm C}.
\end{eqnarray} 
This is the Lyapunov functional of the generalized Smoluchowski-Poisson system.
Indeed, a straightforward calculation yields
\begin{eqnarray}
\label{L4new}
\dot J=\beta\int {1\over\xi\rho}(\nabla p+\rho\nabla\Phi)^{2}d^{3}{\bf r}\ge 0.
\end{eqnarray} 

Let us consider particular cases. For the Boltzmann entropy
(\ref{bol1}), $p=\rho T$, and
\begin{eqnarray}
\label{L5}
J=-\int \rho\ln\rho d^{3}{\bf r}-{1\over 2}\beta\int \rho\Phi d^{3}{\bf r}.
\end{eqnarray} 
Strictly speaking, $\int_{0}^{\rho} p(\rho')/\rho^{'2}d\rho'$ diverges
logarithmically as $\rho'\rightarrow 0$. This means that the general
formula (\ref{L4}) is only marginally correct for an isothermal
equation of state. However, assuming a Boltzmann distribution
(\ref{ge8}) since the begining (see \cite{grand}), we can check
that Eq. (\ref{L5}) is indeed the right formula. Finally, for the Tsallis entropy (\ref{Ts1}),
$p=K\rho^{\gamma}$ with $\gamma=1+1/n$, so that
\begin{eqnarray}
\label{L5h}
J=-\beta n \int p d^{3}{\bf r}-{1\over 2}\beta\int \rho\Phi d^{3}{\bf r}+{\rm C}_{n},
\end{eqnarray} 
where the constant can depend on $n$.  Assuming a Tsallis distribution
(\ref{ge10}) since the begining (see \cite{grand}), we find that
the free energy reads
\begin{eqnarray}
\label{L5ha}
J=-{1\over \gamma-1}\int (\beta K\rho^{\gamma}-\rho)d^{3}{\bf r}-{1\over 2}\beta\int \rho\Phi d^{3}{\bf r}.
\end{eqnarray} 
This expression is consistent with Eq. (\ref{L5h}) and it reduces
to Eq. (\ref{L5}) in the limit $n\rightarrow +\infty$. However, it is
more convenient to write the free energy in the form
\begin{eqnarray}
\label{L5hb}
J=-{\beta K\over \gamma-1}\int (\rho^{\gamma}-\rho)d^{3}{\bf r}-{1\over 2}\beta\int \rho\Phi d^{3}{\bf r},
\end{eqnarray} 
which is also consistent with Eqs. (\ref{L5h}) and (\ref{L5}). Under
this form, we can say that the
high friction limit of a $q$-Tsallis free energy in phase space is a
$\gamma$-Tsallis free energy in configuration space with
$\gamma=(5q-3)/(3q-1)$. We note also that $-J/\beta$ can be written
``${\cal E}-K {\cal S}$'' which is similar to a free energy with $K$
playing the role of a ``polytropic temperature'' (see
\cite{poly}). It is not clear whether this formal result bears more 
physical content than is apparent at first sights.

The generalized Smoluchowski equation (\ref{x11}) can also be written
in the form
\begin{eqnarray}
\label{L6}
{\partial\rho\over\partial t}=\nabla \biggl\lbrack {1\over\xi}(p'(\rho)\nabla\rho+\rho\nabla\Phi)\biggr\rbrack.
\end{eqnarray} 
If we introduce a convex function $C(\rho)$ through the relation
\begin{eqnarray}
\label{L7}
\rho C''(\rho)=\beta p'(\rho),
\end{eqnarray} 
we have equivalently
\begin{eqnarray}
\label{L6new}
{\partial\rho\over\partial t}=\nabla \biggl\lbrack D(\rho C''(\rho)\nabla\rho+\beta\rho \nabla\Phi)\biggr\rbrack,
\end{eqnarray} 
where we have defined  $D=1/\beta\xi$. Noting that Eq. (\ref{L7}) is equivalent to
\begin{eqnarray}
\label{L9}
C(\rho)=\beta\rho\int_{0}^{\rho} {p(\rho')\over \rho'^{2}}d\rho'+C_{1}\rho+C_{2}, 
\end{eqnarray} 
the Lyapunov functional (\ref{L4}) can be rewritten
\begin{eqnarray}
\label{L8}
J=-\int C(\rho)d^{3}{\bf r}-{1\over 2}\beta\int \rho\Phi d^{3}{\bf r}, 
\end{eqnarray} 
where the term proportional to the mass has been droped. The stationary solution of the generalized Smoluchowski equation (\ref{L6new}) is given by
\begin{eqnarray}
\label{es}
C'(\rho)=-\beta\Phi-\alpha. 
\end{eqnarray} 
It can be obtained by maximizing the free energy (\ref{L8}) at fixed mass and temperature. Similarly, the generalized Smoluchowski equation (\ref{L6new}) can be obtained by maximizing the rate of free energy production $\dot J$ at fixed mass and temperature (see Sec. \ref{sec_vrds}).

The case of Boltzmann and Tsallis entropies in configuration space has
been discussed previously. The Fermi-Dirac entropy in configuration
space $S\lbrack \rho\rbrack=-\int \lbrace
\rho\ln\rho+(\rho_{0}-\rho)\ln(\rho_{0}-\rho)\rbrace d^{3}{\bf r}$
leads to an equation of state $p(\rho)=-T\ln (1-\rho/\rho_{0})$. For
$T\rightarrow +\infty$, $p=\rho T$ and for $T\rightarrow 0$, $\rho$ is
a step function with $\rho\sim \rho_{0}$ in the core and $\rho=0$
outside. The corresponding generalized Smoluchowski equation can be
written
\begin{eqnarray}
\label{L6re}
{\partial\rho\over\partial t}=\nabla \biggl\lbrack D'(\nabla\rho+\beta\rho (\rho_{0}-\rho) \nabla\Phi)\biggr\rbrack.
\end{eqnarray}

Coming back to the general case and expliciting the relation between
$\Phi$ and $\rho$, we can rewrite Eq. (\ref{L6new}) in the form
\begin{eqnarray}
\label{L8n}
{\partial\rho\over\partial t}=\nabla \biggl\lbrack D\biggl \lbrace\rho C''(\rho)\nabla\rho+\beta\rho \int \rho({\bf r}',t){\partial u\over\partial {\bf r}}({\bf r}-{\bf r}')d^{3}{\bf r}'\biggr \rbrace\biggr\rbrack.
\end{eqnarray} 
The stationary solution of this equation is determined by the integro-differential equation
\begin{eqnarray}
\label{L9n}
C'(\rho)=-\beta \int \rho({\bf r}')u({\bf r}-{\bf r}')d^{3}{\bf r}'-\alpha.
\end{eqnarray} 
Equation (\ref{L8n}) is one of the most important result of this
paper. The general study of this equation appears to be of
considerable interest in view of its different potential
applications. This project has been initiated by
\cite{crs,sire,anomalous} in particular cases.

\subsection{Generalized stability analysis}
\label{sec_sa}

The equivalence between dynamical and thermodynamical stability for 
the  generalized Smoluchowski equation (\ref{x11}) is proved in 
Appendix \ref{sec_lsK}. In fact, we can go further and
reduce the stability problem to an eigenvalue equation as was done in
the special case of isothermal and polytropic distributions
\cite{crs,anomalous}. Let $\rho$ and $\Phi$ refer to a stationary
solution of Eq. (\ref{x11}) and consider a small perturbation $\delta
\rho$ that conserves mass. We restrict ourselves to spherically
symmetric perturbations (non spherically symmetric perturbations do
not induce instability for non-rotating bodies). Writing $\delta
\rho\sim e^{\lambda t}$ and expanding Eq. (\ref{x11}) to first order,
we find that
\begin{equation}
\lambda\delta \rho={1\over r^{2}}{d\over dr}\biggl \lbrack
{r^{2}\over \xi}\biggl (
{d\delta p\over dr}+\delta \rho {d\Phi\over dr}+
\rho{d\delta\Phi\over dr}\biggr )\biggr\rbrack.
\label{sa1}
\end{equation}
It is convenient to introduce the notation
\begin{equation}
\delta \rho={1\over 4\pi r^{2}}{dq\over dr}.
\label{sa2}
\end{equation}
Physically, $q$ represents the mass perturbation $q(r)\equiv \delta
M(r)=\int_{0}^{r}4\pi {r'}^{2}\delta \rho(r')dr'$ within the sphere of
radius $r$. It satisfies therefore the boundary conditions
$q(0)=q(R)=0$. Substituting Eq.~(\ref{sa2}) in Eq.~(\ref{sa1}) and
integrating, we obtain
\begin{equation}
{\lambda\xi\over r^{2}} q={d\over
dr}\biggl ({p'(\rho)\over r^{2}}{dq\over dr}\biggr )+{1\over r^{2}}{dq\over
dr}{d\Phi\over dr}+4\pi \rho {d\delta\Phi\over dr},
\label{sa3}
\end{equation}
where we have used $q(0)=0$ to eliminate the constant of
integration. Using the condition of hydrostatic equilibrium
$T{d\rho/dr}+\rho{d\Phi/dr}=0$ and the Gauss theorem in perturbed form
${d\delta\Phi/dr}={Gq/r^{2}}$, we can rewrite Eq. (\ref{sa3}) as
\begin{equation}
{\lambda\xi\over 4\pi\rho r^{2}} q={1\over 4\pi \rho}{d\over dr}\biggl ({p'(\rho)\over
r^{2}}{dq\over dr}\biggr )-{1\over 4\pi \rho^{2}}{1\over
r^{2}}{dq\over dr}{dp\over dr}+{Gq\over r^{2}},
\label{sa4}
\end{equation}
or, alternatively,
\begin{equation}
{d\over dr}\biggl ({p'(\rho)\over 4\pi \rho r^{2}}{dq\over dr}\biggr
)+{Gq\over r^{2}}={\lambda\xi\over 4\pi \rho  r^{2}} q,
\label{sa5}
\end{equation}
with $q(0)=q(R)=0$.

Let us now determine the second order variations of the generalized free energy $J$. From Eq. (\ref{L4}), we find that    
\begin{equation}
\delta^{2}J=-\beta\int {p'(\rho)\over 2\rho}(\delta\rho)^{2}d^{3}{\bf r}-{1\over 2}\beta\int \delta\rho\delta\Phi d^{3}{\bf r},
\label{sa6}
\end{equation}
which must be negative (with respect to mass preserving perturbations)
for generalized thermodynamical stability in the canonical
ensemble. Adapting a procedure similar to that followed in
\cite{iso,sire,grand}, we rewrite Eq. (\ref{sa6}) in the form
\begin{equation}
\delta^{2}J=-\beta\int_{0}^{R} {p'(\rho)\over 8\pi\rho r^{2}}\biggl ({dq\over dr}\biggr )^{2} dr-{1\over 2}\beta\int_{0}^{R} {dq\over dr}\delta \Phi dr.
\label{sa7}
\end{equation}
Integrating by parts and using the boundary conditions on $q$, we get
\begin{equation}
\delta^{2}J=\beta\int_{0}^{R} q {d\over dr}\biggl ({p'(\rho)\over 8\pi\rho r^{2}}{dq\over dr}\biggr ) dr+{1\over 2}\beta\int_{0}^{R} q {d\delta \Phi\over dr} dr.
\label{sa8}
\end{equation}
Using the Gauss theorem, we find
\begin{equation}
\delta^{2}J=\beta\int_{0}^{R} q {d\over dr}\biggl ({p'(\rho)\over 8\pi\rho r^{2}}{dq\over dr}\biggr ) dr+{1\over 2}\beta\int_{0}^{R} {Gq^{2}\over r^{2}} dr,
\label{sa9}
\end{equation}
or, equivalently,
\begin{equation}
\delta^{2}J={1\over 2}\beta\int_{0}^{R}dr q\biggl\lbrack  {G\over r^{2}} + {d\over dr}\biggl ({p'(\rho)\over 4\pi\rho r^{2}}{d\over dr}\biggr )\biggr\rbrack q.
\label{sa10}
\end{equation}
The second order variations of free energy will be positive (implying instability) if the differential operator which occurs in the integral has positive eigenvalues.  We need therefore to consider the eigenvalue problem 
\begin{equation}
\biggl\lbrack  {d\over dr}\biggl ({p'(\rho)\over 4\pi\rho r^{2}}{d\over dr}\biggr )+ {G\over r^{2}}\biggr\rbrack q_{\lambda}(r)=\lambda q_{\lambda}(r),
\label{sa11}
\end{equation}
with $ q_{\lambda}(0)=q_{\lambda}(R)=0$. If all the eigenvalues $\lambda$ are negative, then the critical point is a {maximum } of free energy. If at least one eigenvalue is positive, the critical point is an unstable saddle point. The point of marginal stability in the series of equilibria is determined by the condition that the largest eigenvalue is equal to zero $\lambda=0$. We thus have to solve the differential equation
\begin{equation}
{d\over dr}\biggl ({p'(\rho)\over 4\pi\rho r^{2}}{dF\over dr}\biggr )+ {GF(r)\over r^{2}}=0,
\label{sa12}
\end{equation}
with $F(0)=F(R)=0$.

We note that the eigenvalue problems (\ref{sa5}) and (\ref{sa11}) are
similar and that they coincide for marginal stability in continuity
with our previous studies \cite{crs,sire,anomalous}. We have also
found a similar eigenvalue equation by analyzing the stability of
barotropic stars with respect to the Euler-Jeans equations
\cite{iso,poly,grand}. These eigenvalue equations have been solved analytically
(or by using simple graphical constructions) for an isothermal and a
polytropic equation of state in \cite{iso,poly,grand,sire,anomalous}. It is
found that the case of marginal stability ($\lambda=0$) coincides with
the point of minimum generalized temperature $1/\beta$ as predicted
by classical turning point arguments in the canonical ensemble
\cite{katz}. The structure of the perturbation profile that triggers the instability (in particular the number of nodes) has also been determined in our previous papers. The present analysis shows that the structure of the mathematical problem remains the same for a general equation of state  $p=p(\rho)$ even if the solutions cannot be obtained analytically.

\section{Application to stellar dynamics and 2D turbulence}
\label{sec_app}

\subsection{Violent relaxation and metaequilibrium state}
\label{sec_vrms}

We consider a two-dimensional incompressible and inviscid flow
evolving in a plane perpendicular to the direction ${\bf z}$. Let
${\bf u}=-{\bf z}\times \nabla\psi$ denote the velocity field
satisfying the incompressibility condition $\nabla\cdot {\bf
u}=0$. The streamfunction $\psi$ is related to the vorticity $\omega
{\bf z}=\nabla\times {\bf u}$ by the Poisson equation
$\Delta\psi=-\omega$.  More generally, we can consider a relation of
the form $\psi({\bf r})=\int g({\bf r}-{\bf r}')\omega({\bf
r}')d^{2}{\bf r}'$ like, e.g., in the quasi-geostrophic model 
\cite{pedlosky}. We assume that the dynamics is governed by the 2D
Euler equation
\begin{equation}
\label{v3g}
{\partial\omega\over\partial t}+{\bf u}\nabla\omega=0.
\end{equation} 
This equation describes the inviscid evolution of continuous vorticity
flows. It also describes the mean-field evolution of $N\gg 1$ point vortices
before discrete correlations have developed (Vlasov limit)
\cite{kin,houches}. The 2D Euler equation conserves the circulation
\begin{equation}
\label{v2}
\Gamma=\int \omega d^{2}{\bf  r},
\end{equation}
the energy
\begin{equation}
\label{v3}
E={1\over 2}\int \omega\psi d^{2}{\bf  r},
\end{equation} 
and the vorticity moments (equivalent to the Casimirs)
$\Gamma_{n}=\int \omega^{n}d^{2}{\bf r}$. It also conserves the
angular momentum ${L}=\int \omega r^{2} d^{2}{\bf r}$ in a circular
domain and the impulse ${P}=\int \omega y d^{2}{\bf r}$ in a channel
(or in an infinite domain).

When coupled to the Poisson equation, the 2D Euler equation develops
very complex filaments as a result of a mixing process.  In this
sense, the fine-grained vorticity $\omega({\bf r},t)$ never converges
towards a stationary solution. However, if we introduce a
coarse-graining procedure, the coarse-grained vorticity
$\overline{\omega}({\bf r},t)$ rapidly relaxes towards a {\it
metaequilibrium} state. This is called {violent relaxation}, chaotic
mixing or inviscid relaxation. This ``collisionless relaxation'' is
common to all Hamiltonian systems with long range interactions
(stellar systems, point vortices, non-neutral plasmas, HMF
model,...). It physically differs from the ``collisional relaxation''
(or viscous decay) which takes place for much longer times (often
irrelevant). This implies that the order of the limits $t\rightarrow
+\infty$ and $N\rightarrow +\infty$ (or $\nu\rightarrow 0$) is not
interchangeable \cite{houches}.

There has been some attempts to describe the metaequilibrium state in
terms of statistical mechanics \cite{lb,kuzmin,miller,rs}. In the context of
2D hydrodynamics, the statistical metaequilibrium state maximizes the mixing
entropy
\begin{equation}
\label{sm1}
S[\rho]=-\int\rho\ln\rho \ d^{2}{\bf r}d\sigma,
\end{equation}
while conserving circulation, energy and all the Casimirs. The mixing
entropy is the Boltzmann entropy for $\rho({\bf r},\sigma)$,
the density probability of finding the value $\omega=\sigma$ in ${\bf
r}$. The most probable distribution is the Gibbs state
\begin{equation}
\label{sm2}
\rho({\bf r},\sigma)={1\over Z({\bf r})}\chi(\sigma)e^{-(\beta\psi+\alpha)\sigma},
\end{equation}
where  $\chi(\sigma)\equiv {\rm exp}(-\sum_{n>1}\alpha_{n}\sigma^{n})$
accounts for the conservation of the fragile moments
$\Gamma_{n>1}=\int\rho\sigma^{n}d\sigma d^{2}{\bf r}$ and
$\alpha,\beta$ are the usual Lagrange multipliers for $\Gamma$ and $E$
(robust integrals) \cite{grand}. The ``partition function''
$Z=\int\chi(\sigma)e^{-(\beta\psi+\alpha)\sigma}d\sigma$ is determined
by the local normalization condition $\int\rho d \sigma =1$. The equilibrium
coarse-grained vorticity $\overline{\omega}\equiv \int \rho \sigma
d\sigma$ can be expressed as
\begin{equation}
\label{sm4}
\overline{\omega}=-{1\over\beta}{\partial\ln Z\over \partial\psi}=F(\beta\psi+\alpha)=f(\psi).  
\end{equation}
Since $\nabla\overline{\omega}=f'(\psi)\nabla\psi$ and ${\bf u}=-{\bf
z}\times\nabla\psi$, this is a stationary solution of the 2D Euler
equation.  Taking the derivative of Eq. (\ref{sm4}), it is easy to
show that \cite{shallow}:
\begin{equation}
\label{sm5}
\overline{\omega}'(\psi)=-\beta \omega_{2}, \qquad \omega_{2}\equiv \int \rho (\sigma-\overline{\omega})^{2}
d\sigma>0,
\end{equation} 
where $\omega_{2}$ is the centered local enstrophy. We note that the relation $\overline{\omega}=f(\psi)$ is always monotonic, increasing at negative temperatures and decreasing at positive temperatures. Therefore, the coarse-grained vorticity $\overline{\omega}$ extremizes a $H$-function \cite{tremaine}: 
\begin{equation}
\label{v4}
S=-\int C(\omega)d^{2}{\bf r},
\end{equation}
at fixed circulation and energy, where $C(\omega)$ is a convex function, i.e. $C''(\omega)>0$.  Indeed, introducing appropriate Lagrange multipliers and writing the variational principle in the form
\begin{equation}
\label{v5}
\delta S-\beta\delta E-\alpha\delta \Gamma=0,
\end{equation}
we  find that the critical points of entropy at fixed circulation and energy are given by  
\begin{equation}
\label{v6}
C'(\omega)=-\beta\psi-\alpha.
\end{equation}
 The conservation of angular
momentum and impulse can be easily included in the variational
principle (\ref{v5}) by introducing appropriate Lagrange multipliers $\Omega$
and $U$. Equation (\ref{v6}) remains valid provided that $\psi$ is
replaced by the relative streamfunction $\psi'=\psi+{\Omega\over
2}r^{2}-Uy$. Equation (\ref{v6}) can be
written equivalently as
\begin{equation}
\label{v7a}
\omega=F(\beta \psi+\alpha),
\end{equation}
where $F(x)=(C')^{-1}(-x)$. From the identity
\begin{equation}
\label{new}
\omega'(\psi)=-\beta/C''(\omega)
\end{equation}
resulting from Eq. (\ref{v6}), $\omega(\psi)$ is monotonically
decreasing if $\beta>0$ and monotonically increasing if
$\beta<0$. Therefore, for any Gibbs state of the form (\ref{sm4}),
there exists a H-function of the form (\ref{v4}) that the
coarse-grained vorticity $\overline{\omega}$ maximizes (at fixed
$\Gamma$, $E$). It can be shown furthermore that $\overline{\omega}$
maximizes this functional. We note that $C(\omega)$ is a {\it
non-universal} function which depends on the initial conditions. In
general, $S[\overline{\omega}]$ is {\it not} the ordinary Boltzmann entropy
$S_{B}[\omega]=-\int \omega\ln\omega d^{2}{\bf r}$ due to fine-grained
constraints (Casimirs) that modify the form of entropy that we would
naively expect. In the two-levels approximation
$\lbrace \sigma_{0},0\rbrace$ of the statistical theory, $S[\overline{\omega}]$
is the Fermi-Dirac entropy
\begin{equation}
\label{fdV}
S_{F.D.}[\overline{\omega}]=-\int \biggl\lbrace  {\overline{\omega}\over\sigma_{0}}\ln {\overline{\omega}\over\sigma_{0}}+\biggl (1-{\overline{\omega}\over\sigma_{0}}\biggr )\ln \biggl (1-{\overline{\omega}\over\sigma_{0}}\biggr )\biggr \rbrace d^{2}{\bf r},
\end{equation}
leading to the distribution
\begin{equation}
\label{fdV2}
\overline{\omega}={\sigma_{0}\over 1+\lambda  e^{\beta\sigma_{0}\psi}}.
\end{equation}
In the dilute limit $\overline{\omega}\ll\sigma_{0}$, one recovers the
Boltzmann entropy
\begin{equation}
\label{bol1v}
S_{B}[\overline{\omega}]=-\int \overline{\omega}\ln \overline{\omega} d^{2}{\bf r},
\end{equation}
and the isothermal vortex
\begin{equation}
\label{v7}
\overline{\omega}=A e^{-\beta\psi}.
\end{equation}
Other forms of H-functions compatible with the statistical prediction
(\ref{sm4}) are presented in Sec. \ref{sec_class} (from now on, we
drop the bar on $\omega$ except in case of ambiguity).

\subsection{Dynamical stability and thermodynamical analogy}
\label{sec_dsta}

Unfortunately, the statistical theory of violent relaxation is not
very predictive because the initial conditions are not known in
practice and the Casimirs cannot be determined from the coarse-grained
field once the vorticity has mixed (since $\overline{\omega^{n}}\neq
\overline{\omega}^{n}$) \cite{turkington}. In addition, the high order
moments of $\omega$ are altered by non-ideal effects (viscosity,
forcing,...) so that their strict conservation is abusive
\cite{brands2,ellis}. Finally, the relaxation is in general {\it
incomplete} so that the ergodic hypothesis which sustains the
statistical theory is not fulfilled everywhere \cite{staquet,brands}.
One aspect of incomplete violent relaxation is that the
metaequilibrium structures (vortices, galaxies,...) are more confined
than predicted by statistical mechanics.

The only thing that we know for sure is that the metaequilibrium state
reached by the system is a dynamically stable stationary solution of
the 2D Euler-Poisson systems (on a coarse-grained scale). Ellis and
collaborators \cite{ellis} have shown that a strong (nonlinear)
condition of dynamical stability is that ${\omega}$ maximizes a
H-function at fixed circulation and energy. This condition of
dynamical stability can be written
\begin{eqnarray}
\label{v10}
\delta^{2}J=-\int C''(\omega){(\delta \omega)^{2}\over 2}d^{2}{\bf  r}-{1\over 2}\beta\int \delta\omega\delta\psi d^{2}{\bf  r}\le 0,\nonumber\\
\forall\ \delta \omega \mid\ \delta E=\delta \Gamma=0.\qquad\qquad \qquad
\end{eqnarray}
This is {\it similar} to a condition of microcanonical stability in thermodynamics. It is therefore relevant to develop a {\it thermodynamical analogy} and use the same vocabulary as in thermodynamics to analyze the dynamical stability of 2D flows. In this analogy, $S[\omega]$ can be called a generalized entropy, $\beta$ a generalized inverse temperature,... In addition, we can introduce a Legendre transform $J=S-\beta E$ which is similar to a free energy in thermodynamics. The condition that $\omega$ is a maximum of  $J$ at fixed temperature and circulation can be written  
\begin{eqnarray}
\label{v11}
\delta^{2}J=-\int C''(\omega){(\delta \omega)^{2}\over 2}d^{2}{\bf  r}-{1\over 2}\beta\int \delta\omega\delta\psi d^{2}{\bf  r}\le 0,\nonumber\\
\forall\ \delta \omega \mid\ \delta \Gamma=0.\qquad\qquad \qquad
\end{eqnarray}
This is {similar} to a condition of canonical stability in
thermodynamics. The criteria (\ref{v10}) and (\ref{v11}) are not
equivalent if the ``caloric curve'' $\beta(E)$ presents turning points
or bifurcations \cite{oneil,ellis}. This corresponds to a situation of
ensemble inequivalence in thermodynamics. Since canonical stability
implies microcanonical stability (but not the converse),
the stability criterion (\ref{v10}) is stronger than (\ref{v11}). This
has important implications in geophysical and jovian fluid dynamics \cite{ellis,bthese}.

The H-function $S[\omega]$ maximized by the system at metaequilibrium
is {\it non-universal}. It depends on the initial conditions {\it and}
on the strength of mixing. The Tsallis entropy
\begin{equation}
\label{Ts2}
S_{q}[\omega]=-{1\over q-1}\int (\omega^{q}-\omega)  d^{2}{\bf r},
\end{equation}
is a particular H-function that {\it sometimes} occurs in 2D hydrodynamics
\cite{boghosian}. However, this is essentially fortuitous, as
discussed in \cite{brands}. In addition, in this context, Tsallis
functional is not a true entropy. Its maximization 
at fixed circulation and energy is a condition of (nonlinear) dynamical
stability not a condition of thermodynamical stability. It leads to a
{\it particular} class of stable stationary solutions of the 2D Euler
equation characterized by
\begin{equation}
\label{v9}
\omega=A(\lambda-\psi)^{n},
\end{equation} 
with $A=\lbrack (q-1)\beta/q\rbrack^{1\over q-1}$ and $\lambda=\lbrack
1-(q-1)\alpha\rbrack /(q-1)\beta$. We shall call this class of
vortices {\it polytropic vortices} by analogy with stellar polytropes
\cite{bt}. The index $n$ of the polytrope is related to the parameter
$q$ by the relation $n=1/(q-1)$. Isothermal vortices are recovered in
the limit $q\rightarrow 1$ (i.e. $n\rightarrow +\infty$). For $q=2$,
i.e. $n=1$, the relationship between $\omega$ and $\psi$ is linear
\cite{boghosian}. 

To conclude this section, we note that dynamical stability results
similar to (\ref{v10}) and (\ref{v11}) have been obtained for the
Vlasov-Poisson system \cite{ipser,grand}. Therefore, the generalized
thermodynamical formalism developed in Sec. \ref{sec_gt} can be used
to study the (nonlinear) dynamical stability of collisionless stellar
systems in astrophysics and other systems with long-range interactions
described by the Vlasov equation.

\subsection{Relaxation equations for 2D flows}
\label{sec_vrds}

According to the previous discussion, an important problem in 2D
hydrodynamics is to construct particular stationary solutions of the
2D Euler equation with strong stability properties. We shall consider
solutions that maximize a H-function $S[\omega]$ at fixed circulation
$\Gamma$ and energy $E$ \cite{ellis}. The explicit construction of
such solutions is non-trivial. Expoiting the thermodynamical analogy
discussed previously and using the MEPP, we can obtain relaxation
equations (similar to Fokker-Planck equations) that can be used as
numerical algorithms to construct arbitrary stable stationary solutions of the
2D Euler equation.  To apply the MEPP, we write the relaxation
equation in the form
\begin{equation}
\label{g2}
{\partial \omega\over\partial t}+{\bf u}\cdot \nabla \omega=-\nabla\cdot {\bf J}_{\omega},
\end{equation} 
where the diffusion current ${\bf J}_{\omega}$ has to be determined.
The form of Eq. (\ref{g2}) ensures the conservation of circulation
provided that ${\bf J}_{\omega}\cdot {\bf n}=0$ on the domain boundary
(with normal vector ${\bf n}$). From Eqs. (\ref{v3}), (\ref{v4}) and
(\ref{g2}), it is easy to put the time variations of energy and entropy
in the form
\begin{equation}
\label{g3}
\dot E=\int {\bf J}_{\omega}\cdot \nabla\psi d^{2}{\bf r},
\end{equation}
\begin{equation}
\label{g4}
\dot S=-\int C''(\omega) {\bf J}_{\omega}\cdot \nabla\omega d^{2}{\bf r},
\end{equation}
where we have used straightforward integrations by parts. Following the MEPP, we now determine the optimal current ${\bf J}_{\omega}$ which maximizes the rate of entropy production (\ref{g4}) while satisfying the conservation of energy $\dot E=0$ and the constraint
\begin{equation}
\label{g5}
{J_{\omega}^{2}}\le C({\bf r},t).
\end{equation}
This maximization problem  leads to the optimal current
\begin{equation}
\label{g6}
{\bf J}_{\omega}=-D\biggl\lbrack \omega C''(\omega)\nabla\omega+\beta(t)\omega\nabla\psi\biggr \rbrack.
\end{equation}
The time evolution of the Lagrange multiplier $\beta(t)$ is determined
by introducing Eq. (\ref{g6}) in the energy constraint (\ref{g3}),
using $\dot E=0$. This yields
\begin{equation}
\label{g7}
\beta(t)=-{\int D \omega C''(\omega)\nabla\omega\cdot \nabla\psi d^{2}{\bf r}\over \int D \omega (\nabla\psi)^{2} d^{2}{\bf r}}.  
\end{equation}
Introducing the optimal current (\ref{g6}) in Eq. (\ref{g2}), we obtain a relaxation equation of the form
\begin{equation}
\label{g8}
{\partial \omega\over\partial t}+{\bf u}\cdot \nabla \omega=\nabla\biggl\lbrace D\biggl\lbrack \omega C''(\omega)\nabla\omega+\beta(t)\omega\nabla\psi\biggr\rbrack\biggr\rbrace.
\end{equation}  
The first term is a generalized diffusion and the
second term is a drift. The function $\beta(t)$ can be considered as a
time dependant inverse temperature (possibly negative). It evolves
with time so as to conserve the total energy $E$ (microcanonical
description). The drift coefficient $\xi=D\beta$ is a generalized
Einstein relation.  We shall use Eq. (\ref{g8}) when $\omega\ge
0$. When $\omega$ can be positive and negative, we shall prefer the
alternative form
\begin{equation}
\label{g9}
{\partial \omega\over\partial t}+{\bf u}\cdot \nabla \omega=\nabla\biggl\lbrace D'\biggl\lbrack \nabla\omega+{\beta(t)\over C''(\omega)}\nabla\psi\biggr\rbrack\biggr\rbrace,
\end{equation}
Equation (\ref{g9}) is obtained from Eq. (\ref{g8}) by setting $D'=D\omega C''(\omega)$. It is straightforward to check that Eq. (\ref{g8}) with the constraint (\ref{g7}) satisfies a $H$-theorem for the generalized entropy (\ref{v4}). Indeed, Eq. (\ref{g4}), (\ref{g3}) and (\ref{g6}) lead to 
\begin{equation}
\label{g10}
\dot S=\int {J_{\omega}^{2}\over D\omega}d^{2}{\bf r},
\end{equation}
where we have used $\dot E=0$. If $\omega\ge 0$, then Eq. (\ref{g10}) is positive provided that $D\ge 0$. If we use the alternative equation  (\ref{g9}), we have $D\omega=D'/C''(\omega)$ so that Eq. (\ref{g10}) is positive whatever the sign of $\omega$ provided that $D'\ge 0$.  At equilibrium $\dot S=0$, hence ${\bf
J}_{\omega}={\bf 0}$, which is equivalent to
\begin{equation}
\label{g11}
\nabla C'(\omega)+\beta\nabla\psi={\bf 0}.
\end{equation}
Integrating, we get
\begin{equation}
\label{g12}
C'(\omega)=-\beta \psi-\alpha.
\end{equation}
which returns Eq. (\ref{v6}) with $\beta=\lim_{t\rightarrow
+\infty}\beta(t)$. Therefore, as expected, a stationary solution of
Eq. (\ref{g8}) extremizes the generalized entropy (\ref{v4}) at fixed
energy and circulation. In addition, it is shown in Appendix
\ref{sec_lsV} that the relation (\ref{l15}) remains valid so that only
{\it maxima} of $S$ at fixed $E$ and $\Gamma$ are selected by the relaxation
equation.  The relaxation equation appropriate to the canonical
situation is obtained by maximizing $\dot J=\dot S-\beta\dot E$ with
the constraint (\ref{g5}). This leads again to an optimal current of
the form (\ref{g6}) but with constant $\beta$. In addition,
Eqs. (\ref{g10}) and (\ref{l15}) remain valid with $J$ in place of
$S$. Hence, the free energy $J$ increases monotonically until a state
of maximum free energy is reached.

To conclude this section, we shall discuss particular cases. For the Boltzmann entropy, $C''(\omega)=1/\omega$ and Eq. (\ref{g8}) reduces to
\begin{equation}
\label{gq1}
{\partial \omega\over\partial t}+{\bf u}\cdot \nabla \omega=\nabla\biggl\lbrack D\biggl (\nabla\omega+\beta\omega \nabla\psi\biggr )\biggr\rbrack.
\end{equation}
For the Fermi-Dirac entropy, $C''(\omega)=1/\omega(\sigma_{0}-\omega)$ and Eq. (\ref{g9}) yields
\begin{equation}
\label{gq2}
{\partial \omega\over\partial t}+{\bf u}\cdot \nabla \omega=\nabla\biggl\lbrace D'\biggl\lbrack \nabla\omega+\beta\omega(\sigma_{0}-\omega) \nabla\psi\biggr\rbrack\biggr\rbrace.
\end{equation}
For the
Tsallis entropy, $C''(\omega)=q \omega^{q-2}$ and Eq. (\ref{g8}) becomes
\begin{equation}
\label{g14}
{\partial \omega\over\partial t}+{\bf u}\cdot \nabla \omega=\nabla\biggl\lbrace D\biggl\lbrack \nabla\omega^{q}+\beta\omega \nabla\psi\biggr\rbrack\biggr\rbrace.
\end{equation}

\subsection{A physical numerical algorithm}
\label{sec_na}

Due to the thermodynamical analogy, the relaxation
equations proposed in Sec. \ref{sec_vrds} can provide a powerful
numerical algorithm to compute arbitrary dynamically stable stationary
solutions of the 2D Euler equation. Since we are only interested by the
stationary solution, we can forget the advective term in
Eq. (\ref{g9}) and fix $D$ to an arbitrary positive constant. We
propose therefore the physical numerical algorithm
\begin{equation}
\label{ax1}
{\partial {\omega}\over\partial t}=\nabla\biggl\lbrace D\biggl\lbrack \nabla{\omega}+{\beta(t)\over C''({\omega})}\nabla\psi\biggr\rbrack\biggr\rbrace,
\end{equation}
\begin{equation}
\label{ax2}
\beta(t)=-{\int D\nabla{\omega}\nabla\psi d^{2}{\bf r}\over \int D{(\nabla\psi)^{2}\over C''({\omega})}d^{2}{\bf r}}.
\end{equation}
These equations satisfy the conservation of circulation and energy
(robust integrals) and increase the generalized entropy (\ref{v4})
until the system has reached a {\it maximum} of $S$ at fixed $\Gamma$
and $E$. We have seen indeed that a minimum or a saddle point of
$S[\omega]$ are linearly unstable via Eqs. (\ref{ax1})-(\ref{ax2}).
Note that specifying $C(\omega)$ does not directly determine the
equilibrium state because many bifurcations can occur in parameter
space $(\Gamma,E)$. There can also exist {\it local} entropy maxima
(similar to metastable states in thermodynamics) leading to a
complicated notion of {\it basin of attraction} (see \cite{crs} in a
related context). These equations could be used to study dynamical
stability problems on a new angle \cite{ellis}. Indeed, if a
stationary solution of Eq.  (\ref{ax1}) is stable at fixed inverse
temperature $\beta$ or at fixed energy $E$, then it is nonlinearly
stable with respect to the 2D Euler equation. On the other hand, if a
stationary solution of Eq.  (\ref{ax1}) is stable at fixed energy $E$
(variable $\beta(t)$) but not at fixed inverse temperature $\beta$,
then it violates Arnold's sufficient conditions of stability (see
Appendix \ref{sec_suff}) but is, however, dynamically stable with
respect to the 2D Euler equation. The relaxation equations proposed in
Sec. \ref{sec_gt} could be used similarly as numerical algorithms to
construct nonlinearly dynamically stable stationary solutions of the
Vlasov equation.

\subsection{A simplified parametrization of 2D turbulence}
\label{sec_relax}

The generalized Fokker-Planck equations derived in Sec. \ref{sec_vrds}
can also provide a simplified parametrization of 2D turbulence. The
thermodynamical parametrization proposed by Robert \& Sommeria
\cite{rsmepp} can be written
\begin{equation}
\label{rob1}
{\partial \rho\over\partial t}+{\bf u}\cdot \nabla \rho=\nabla\biggl\lbrace D\biggl\lbrack \nabla\rho+\beta(t)\rho(\sigma-\rho)\nabla\psi\biggr\rbrack\biggr\rbrace,
\end{equation}
where $\rho({\bf r},\sigma,t)$ denotes the density probability of
finding the vorticity level $\sigma$ in ${\bf r}$ at time
$t$. Equation (\ref{rob1}) incorporates a turbulent viscosity $D$ and an
additional term interpreted as a {\it systematic drift}
\cite{drift}. The drift is due to the inhomogeneity of the medium and
is supported by arguments of kinetic theory in simplified models
(point vortices, quasilinear approximation,...)
\cite{drift,kin,quasiE}. Usual parametrizations including a single
turbulent viscosity correspond to the infinite temperature limit
($\beta=0$) of the thermodynamical parametrization. Equation
(\ref{rob1}) increases the mixing entropy $S[\rho]$ while conserving
the energy and all the Casimirs. For $t\rightarrow +\infty$, the
solution converges to the Gibbs state (\ref{sm2}).

The equations of Robert \& Sommeria \cite{rsmepp} are complicated
because they take into account the conservation of {\it all} the
Casimirs. This clearly leads to practical difficulties. This also leads to
physical difficulties because the strict conservation of all the Casimirs
is abusive as discussed in \ref{sec_vrds}. We could try to simplify
the problem by writing a hierarchy of equations for the moments of
$\rho$. The first equation of this hierarchy is
\begin{equation}
\label{rob2}
{\partial \overline{\omega}\over\partial t}+{\bf u}\cdot \nabla \overline{\omega}=\nabla\biggl\lbrace D\biggl\lbrack \nabla\overline{\omega}+\beta(t)\omega_{2}\nabla\psi\biggr\rbrack\biggr\rbrace,
\end{equation}
where $\omega_{2}$ is the local centered enstrophy defined in
Eq. (\ref{sm5}).  However, we are now led to a difficult closure
problem. Kazantsev {\it et al.} \cite{kazantsev} have proposed to
close the hierarchy of equations by a Gaussian approximation. This
leads to an equilibrium state corresponding to a minimum enstrophy
state. Although this state may be relevant in some particular oceanic
problems (Fofonoff flows), this is not expected to be general.

In this paper, we propose to close the hierarchy of
moment equations by a relation of the form 
\begin{equation}
\label{impo}
\omega_{2}={1\over C''(\omega)},
\end{equation}
which can be deduced from Eqs. (\ref{sm5}) and (\ref{new}). This
relation is valid at statistical equilibrium but we shall use it out
of equilibrium as a convenient approximation. In a sense, this
supposes that the Lagrange multipliers $\alpha_{n}$ associated to the
fragile constraints $\Gamma_{n>1}$ in the Gibbs state (\ref{sm2}) can
be treated canonically \cite{ellis}.  Substituting Eq. (\ref{impo}) in
Eq. (\ref{rob2}) we obtain the simplified parametrization
\begin{equation}
\label{ax1a}
{\partial {\omega}\over\partial t}+{\bf u}\cdot \nabla {\omega}=\nabla\biggl\lbrace D\biggl\lbrack \nabla{\omega}+{\beta(t)\over C''({\omega})}\nabla\psi\biggr\rbrack\biggr\rbrace,
\end{equation}
\begin{equation}
\label{ax2b}
\beta(t)=-{\int D\nabla{\omega}\nabla\psi d^{2}{\bf r}\over \int D{(\nabla\psi)^{2}\over C''({\omega})}d^{2}{\bf r}},
\end{equation}
which coincides with Eqs. (\ref{g9}) and (\ref{g7}).  The function
$C(\omega)$ is a {\it non-universal} function which encapsulates the
complexity of the fine-grained dynamics and which depends on the
situation contemplated (ocean dynamics, jovian atmosphere, decaying 2D
turbulence...). For a given physical situation, we propose to select a
form of $C(\omega)$ {\it a priori} and compute the corresponding
equilibrium state. Then, we can check {\it a posteriori} whether it
was a good choice by comparing the result with the information that we
have on the system. This is similar to the notion of {\it prior
vorticity distribution} proposed in \cite{ellis}. Of course, this
approach is essentially phenomenological and explanatory but it allows
to deal with complex situations which were previously inaccessible.

To obtain an operational sub-grid scale parametrization of 2D
turbulence, it remains for one to specify the value of the diffusion
coefficient. Heuristic arguments \cite{rr,csr} or more formal kinetic
theory \cite{quasiE} suggest that $D=K\epsilon^{2}\omega_{2}^{1/2}$
where $K$ is a constant of order unity and $\epsilon$ is the scale of
unresolved fluctuations. This formula is a relatively
direct consequence of the general Taylor expression of the turbulent
viscosity \cite{taylor} and it proved to be relevant in oceanic modelling
\cite{kazantsev}. Furthermore, it has been shown in previous works
\cite{rr,csr} that the spatial dependance of the diffusion coefficient
is important to take into account the problem of {\it incomplete
relaxation} and the formation of self-confined vortices
\cite{jfm2}. With the closure relation (\ref{impo}), the diffusion 
coefficient can be expressed in terms of $\omega$ as
\begin{equation}
\label{ax2bb}
D={K\epsilon^{2}\over \sqrt{C''(\omega)}}.
\end{equation}
The same simplifications could be introduced in the parametrization of
the gravitational Vlasov-Poisson proposed in \cite{csr}.

\subsection{Classification of generalized entropies}
\label{sec_class}

Our parametrization involves a free function $C(\omega)$. This
indetermination is intrinsic to the problem of 2D turbulence and not
really a flaw of our approach: there is {\it no} universal entropy
$S[\omega]$.  In order to reduce this indetermination, we shall argue
that generalized entropies can be regrouped in ``classes of
equivalence'' with the underlying idea that functionals of the same
class will produce similar results. Then, for a given physical
situation, it will be possible to pick a form of $C(\omega)$ in the
corresponding class of equivalence and use it in the parametrization (\ref{ax1a})-(\ref{ax2bb}). 

We shall now present typical forms of ``generalized entropies''
$S[\omega]$ that appeared in the literature.  Specifically, we shall
prescribe  analytical forms of the vorticity distribution
$\chi(\sigma)$ in the Gibbs state (\ref{sm2}) and compute the
corresponding $C(\omega)$.  These analytical expressions can be
considered as prototypical examples of more realistic
distributions. Consider first the case where the fine-grained
vorticity takes only two values $\sigma_{0}$ and
$\sigma_{1}>\sigma_{0}$ so that
$\chi(\sigma)=\chi_{0}\delta(\sigma-\sigma_{0})+\chi_{1}\delta(\sigma-\sigma_{1})$. In
that case, $\omega(\psi)$ is the distribution
\begin{equation}
\label{c3}
\omega=\sigma_{0}+{\sigma_{1}-\sigma_{0}\over 1+e^{(\sigma_{1}-\sigma_{0})(\beta\psi+\alpha)}},
\end{equation}
which can also be written as a $\tanh$,
\begin{equation}
\label{c4}
\omega={\sigma_{0}+\sigma_{1}\over 2}-{\sigma_{1}-\sigma_{0}\over 2}\tanh\biggl\lbrack {\sigma_{1}-\sigma_{0}\over 2}(\beta\psi+\alpha)\biggr\rbrack.
\end{equation}
For this distribution, $\omega\rightarrow \sigma_{1},\sigma_{0}$ when
$-\beta\psi\rightarrow +\infty,-\infty$. Using Eq. (\ref{v6}), we get
\begin{equation}
\label{c1}
C(\omega)=p\ln p+(1-p)\ln (1-p),\quad \omega=p\sigma_{1}+(1-p)\sigma_{0}.
\end{equation} 
This implies
\begin{equation}
\label{c2}
C'(\omega)={1\over\sigma_{1}-\sigma_{0}}\ln\biggl ({\omega-\sigma_{0}\over\sigma_{1}-\omega}\biggr ), \qquad C''(\omega)={1\over (\omega-\sigma_{0})(\sigma_{1}-\omega)}.
\end{equation}
Substituting Eq. (\ref{c2}) in the
general parametrization (\ref{ax1a}), we obtain
\begin{equation}
\label{c5}
{\partial {\omega}\over\partial t}+{\bf u}\cdot \nabla {\omega}=\nabla\biggl\lbrace D\biggl\lbrack \nabla{\omega}+\beta(t)(\omega-\sigma_{0})(\sigma_{1}-\omega)\nabla\psi\biggr\rbrack\biggr\rbrace,
\end{equation}
with a diffusion coefficient 
\begin{equation}
\label{c6}
D=K\epsilon^{2}\sqrt{(\omega-\sigma_{0})(\sigma_{1}-\omega)}.
\end{equation}
These equations coincide with the two-levels approximation of the
thermodynamical parametrization (\ref{rob1}). This
two-levels approximation has been used in relation with shear layer
instability \cite{staquet}, vortex merging \cite{rsmepp} and in a
model of Jupiter's great red spot \cite{bs}.

Turkington \cite{turkington} has considered the distribution $\chi(\sigma)=\chi$ for $\sigma_{0}\le \sigma\le \sigma_{1}$ and $\chi(\sigma)=0$ otherwise. The choice $\alpha_{n}=0$ for $n>1$ in Eq. (\ref{sm2}) amounts to maximizing the entropy (\ref{sm1})
at fixed energy and circulation, neglecting the higher order vorticity
moments.  If we assume for simplicity that $\sigma_{0}=-\lambda$ and
$\sigma_{1}=+\lambda$ then $\omega(\psi)$ is the Langevin function 
\begin{equation}
\label{c7}
\omega=\lambda L\lbrack -\lambda(\beta\psi+\alpha)\rbrack, \qquad L(x)=\coth(x)-{1\over x}.
\end{equation}
It does not seem possible to write down the function $C(\omega)$
explicitly. However, the relationship (\ref{c7}) is {qualitatively}
similar to (\ref{c4}) so we argue that the Langevin-type model falls
in the same ``class of equivalence'' as the Fermi-Dirac-type model (i.e.,
they will produce the same type of equilibrium states and bifurcations).

Consider now a situation in which the fine-grained vorticity can take
three values $\sigma_{1}$, $\sigma_{2}$ and $0$ so that
$\chi(\sigma)=\delta(\sigma)+\chi_{1}\delta(\sigma-\sigma_{1})+\chi_{2}\delta(\sigma-\sigma_{2})$. Consider
furthermore the dilute limit of the statistical theory in which
$Z\simeq 1$. In that case, the probability of the non-zero levels is
given by $p_{i}=A_{i}{\rm exp}\lbrack
-(\beta\psi+\alpha)\sigma_{i}\rbrack$ and
$\omega=p_{1}\sigma_{1}+p_{2}\sigma_{2}$. This dilute limit
corresponds to the point vortex model 
\cite{jm}. If we assume furthermore, for simplicity, that
$\sigma_{0}=-\lambda$ and $\sigma_{1}=+\lambda$ we get the sinh-Poisson
relation
\begin{equation}
\label{c9}
\omega=-2A\lambda\sinh(\lambda(\beta\psi+\alpha)).
\end{equation}
For this distribution, $\omega\rightarrow \pm\infty$ as
$-\beta\psi\rightarrow \pm\infty$. The sinh-relationship is observed
in the late stages of 2D turbulence when the initial condition is a
random vorticity field \cite{mont}.  This is because the vortices that
form at intermediate times are very intense and isolated as in a point
vortex gas. Note that the statistical approach based on the
conservation of all the Casimirs does not work in that case because of
viscous effects
\cite{brands2}.  Using Eq. (\ref{v6}), we
find that
\begin{equation}
\label{c10}
C'(\omega)={1\over\lambda} \sinh^{-1}\biggl ({\omega\over 2A\lambda}\biggr ).
\end{equation}
Then,
\begin{equation}
\label{c11}
C(\omega)={\omega\over\lambda} \sinh^{-1}\biggl ({\omega\over 2A\lambda}\biggr )-{1\over\lambda}\sqrt{4A^{2}\lambda^{2}+{\omega^{2}}},
\end{equation}
and 
\begin{equation}
\label{c12}
C''(\omega)={1\over \lambda}{1\over \sqrt{4A^{2}\lambda^{2}+\omega^{2}}}.
\end{equation}
Therefore, in the case of 2D decaying turbulence with  random initial conditions, the parametrization (\ref{ax1a}) that we propose is
\begin{equation}
\label{c13}
{\partial {\omega}\over\partial t}+{\bf u}\cdot \nabla {\omega}=\nabla\biggl\lbrace D\biggl\lbrack \nabla{\omega}+\beta(t)\lambda\sqrt{4A^{2}\lambda^{2}+\omega^{2}}\nabla\psi\biggr\rbrack\biggr\rbrace,
\end{equation}
with a diffusion coefficient 
\begin{equation}
\label{c14}
D=K\epsilon^{2}\sqrt{\lambda}(4A^{2}\lambda^{2}+\omega^{2})^{1/4}.
\end{equation}
If we assumes a Poissonian weight factor $\chi(\sigma)={\rm
exp}(-|\sigma|/q)$ in Eq. (\ref{sm2}), we get 
\begin{equation}
\label{pas}
\omega=-{2q^{2}(\beta\psi+\alpha)\over 1-q^{2}(\beta\psi+\alpha)^{2}}.
\end{equation}
Pasmanter \cite{pasmanter} has noted that this relationship yields
results similar to those obtained with the sinh-relationship, so they fall
in the same ``class of equivalence''.

If we now assume that the local distribution of vorticity is Gaussian so that $\chi(\sigma)=e^{-\alpha_{2}\sigma^{2}}$ in Eq. (\ref{sm2}), then the $\omega-\psi$ relationship is linear \cite{miller} and can be written
\begin{equation}
\label{c15}
\omega=-\Omega_{2}(\beta\psi+\alpha),
\end{equation}
where $\Omega_{2}=\omega_{2}$ is a constant equal to the centered
variance of the vorticity distribution. Such a linear relationship
sometimes occurs in geophysics \cite{fofonoff}. The
choice $\alpha_{n}=0$ for $n>2$ is equivalent to maximizing the
entropy (\ref{sm1}) at fixed energy, circulation and enstrophy
(neglecting the higher order moments). The function $C(\omega)$ is
\begin{equation}
\label{c16}
C(\omega)={1\over 2\Omega_{2}}\omega^{2},
\end{equation}
so that the generalized entropy $S[\omega]$ is proportional to minus
the (coarse-grained) enstrophy $\Gamma_{2}=\int \overline{\omega}^{2}d^{2}{\bf
r}$. Therefore, Eq. (\ref{c15}) can also be obtained by minimizing the
enstrophy at fixed energy and circulation \cite{leith}. A linear
$\omega-\psi$ relationship is also obtained in the {\it strong mixing
limit} of the statistical theory providing an inviscid justification
of the minimum enstrophy principle \cite{jfm1}.

Another choice of probability distribution has been proposed
by Ellis {\it et al.} \cite{ellis} in a model of jovian atmosphere where
the  skewness is expected to play a crucial role. They relate
$\chi(\sigma)$ in the general formula (\ref{sm2}) to the gamma density
$z^{a-1}e^{-z} \ (z\le 0)$ and find that
\begin{equation}
\label{c17}
\omega={-\Omega_{2}(\beta\psi+\alpha)\over 1+\lambda\Omega_{2}(\beta\psi+\alpha)},
\end{equation}
where $\Omega_{2}$ is equal to the variance of $\chi(\sigma)$ and
$2\lambda\Omega_{2}^{1/2}$ is equal to the skewness of $\chi(\sigma)$.
For this relationship, $\omega\rightarrow -\infty$ as
$-(\beta\psi+\alpha)\rightarrow 1/\lambda\Omega_{2}$ and
$\omega\rightarrow -1/\lambda$ as $-\beta\psi\rightarrow
+\infty$. Therefore, the gamma model (\ref{c17}) is somewhat
intermediate between the tanh model (\ref{c4}) and the sinh model
(\ref{c9}).  Using Eq. (\ref{v6}), we find that the corresponding
generalized entropy is
\begin{equation}
\label{c18}
C(\omega)={1\over\lambda\Omega_{2}}\biggl\lbrack \omega-{1\over\lambda}\ln (1+\lambda{\omega})\biggr\rbrack.
\end{equation}
This yields
\begin{equation}
\label{c19}
C'(\omega)={\omega\over \Omega_{2}(1+\lambda\omega)},\qquad C''(\omega)={1\over \Omega_{2}(1+\lambda\omega)^{2}}. 
\end{equation}
For $\lambda\rightarrow 0$, we recover the gaussian model (\ref{c15})
as a special case of the gamma model. The parametrization (\ref{ax1a})
that we propose for this model is
\begin{equation}
\label{c20}
{\partial {\omega}\over\partial t}+{\bf u}\cdot \nabla {\omega}=\nabla\biggl\lbrace D\biggl\lbrack \nabla{\omega}+\beta(t)\Omega_{2}(1+\lambda\omega)^{2}\nabla\psi\biggr\rbrack\biggr\rbrace,
\end{equation}
with a diffusion coefficient  
\begin{equation}
\label{c21}
D=K\epsilon^{2}\Omega_{2}^{1/2}(1+\lambda\omega).
\end{equation}
For $\lambda\rightarrow 0$ (gaussian approximation), we recover the
first moment equation of the parametrization used by Kazantsev {\it et
al.} \cite{kazantsev} in a barotropic ocean model.

\section{Conclusion}
\label{sec_conclusion}

In this paper, we have introduced a new class of relaxation equations
associated with a generalized thermodynamical framework. The
Fokker-Planck and nonlinear Fokker-Planck equations corresponding to
Boltzmann and Tsallis entropies are recovered as a special case. The
potential applications of this new class of relaxation equations is
considerable and concerns various domains of physics such as stellar
dynamics, 2D turbulence, plasma physics, chemotaxis, porous media
etc...  For example, these equations can serve as numerical algorithms
to compute arbitrary nonlinearly dynamically stable stationary
solutions of Vlasov-Poisson or 2D Euler-Poisson systems. Physical
applications of these equations and numerical simulations will be
presented in future works. 

We have also clarified the concept of generalized thermodynamics
introduced by Tsallis \cite{tsallis} and applied to self-gravitating
systems and two-dimensional vortices by Plastino \& Plastino
\cite{plastino} and Boghosian \cite{boghosian}. If we consider a
collection of stars or point vortices and take the long time limit
($t\rightarrow +\infty$ at fixed $N\gg 1$), the statistical
equilibrium state resulting from a ``collisional'' evolution is
correctly described by the ordinary Boltzmann entropy $S_{B}[f]=-\int
f\ln f d^{3}{\bf r}d^{3}{\bf v}$ or
$S_{B}[\omega]=-\int\omega\ln\omega d^{2}{\bf r}$ but the
thermodynamic limit is unusual (and does {\it not} correspond to
$N,V\rightarrow +\infty$ with $N/V$ fixed)
\cite{houches}.  In the case of stellar systems, there is no true
equilibrium state  but this corresponds to
important physical processes (evaporation, gravothermal catastrophe)
not to a breakup of thermodynamics \cite{grand}. The kinetic theory of stars and point vortices is extremely
complicated due to the long-range nature of the interactions. Kinetic
equations can be derived rigorously by using projection operator
technics but they are non-Markovian and integro-differential
\cite{kandrup,kin}. The $H$-theorem for the Boltzmann entropy cannot 
be proved without further approximations.  In addition, the diffusion
coefficient depends on position, velocity and time and this can lead
to a confinement of the structure in physical or phase space.  These
complicated effects (space and time delocalizations) can induce, for
intermediate collisional times, a deviation with respect to an ideal
statistical evolution. However, the infinite time limit should be
described by the Boltzmann distribution. 

On the other hand, for times $t\ll t_{relax}$ and $N\rightarrow
+\infty$, a collection of stars or point vortices (or continuous
vorticity flows) can achieve a metaequilibrium state as a result of a
``collisionless'' (or inviscid) violent relaxation. Note that the
domain of validity of the collisionless regime is huge for such
systems since the collisional relaxation time $t_{relax}\sim
{N\over\ln N}t_{D}$ is much larger than the dynamical time $t_{D}$
\cite{houches}.  The correct entropy is the Boltzmann entropy
$S[\rho]=-\int\rho\ln\rho d\eta d^{3}{\bf r}d^{3}{\bf v}$ or
$S[\rho]=-\int\rho\ln\rho d\sigma d^{2}{\bf r}$ for $\rho$, the local
distribution of phase levels or vorticity levels. The statistical
metaequilibrium state is obtained by maximizing the Boltzmann entropy
$S[\rho]$ while conserving energy, circulation and an infinite class
of invariants called the Casimirs. Alternatively, the coarse-grained
distribution function or coarse-grained vorticity can also be obtained
by maximizing a H-function $S[\overline{f}]=-\int
C(\overline{f}) d^{3}{\bf r}d^{3}{\bf v}$ or
$S[\overline{\omega}]=-\int C(\overline{\omega}) d^{2}{\bf r}$, where
$C$ is a convex function, while conserving only mass (circulation) and
energy (robust constraints).  This functional depends on the initial
conditions and is therefore non-universal.  In this context,
``generalized entropies'' arise due to the presence of fine-grained
constraints (Casimirs) that modify the macroscopic form of entropy
that we would naively expect. This is a particularity of {\it
continuous Hamiltonian systems} and this makes the equilibrium state
difficult (if not impossible) to predict. A classification of generalized
entropies in {\it classes of equivalence} can however be attempted.

Furthermore, it can happen that the mixing process during the
collisionless relaxation is not sufficient to justify the ergodic
hypothesis which sustains the statistical theory. In that case, the
metaequilibrium state results from an {\it incomplete} violent
relaxation. One possibility to take into account incomplete mixing and
non-ergodicity is to introduce additional kinetic constraints in the
statistical approach. This can be done by using relaxation equations
with a variable diffusivity related to the local fluctuations of the
vorticity or distribution function \cite{rr,csr}. Accordingly, the
physical picture that emerges is the following: during violent
relaxation, the system has the {\it tendency} to reach the most mixed
state described by the Gibbs distribution. However, as it approaches
equilibrium, the mixing becomes less and less efficient and the system
settles on a stationary state which is not the most mixed state.
Then, the evolution is stopped until other effects (collisions,
viscosity,...) come into play. The state resulting from an incomplete
violent relaxation is a (nonlinearly) dynamically stable stationary
solution of the Vlasov or 2D Euler equation on the coarse-grained
scale. A strong condition of stability is that it maximizes a
$H$-function at fixed mass (circulation) and energy
\cite{ipser,ellis}. Since this condition of nonlinear dynamical
stability is {\it similar} to a condition of thermodynamical
stability, we can use a thermodynamical analogy to study the dynamical
stability of stellar systems and 2D vortices. In this analogy, the
H-function can be regarded as a generalized entropy. The H-function
maximized by the system at metaequilibrium depends on the initial
conditions, on the efficiency of mixing and on non-ideal
effects. Tsallis entropy is just a particular $H$-function leading to
polytropic distributions. It can sometimes provide a good fit of the
metaequilibrium state in case of incomplete relaxation
\cite{boghosian} but this is not general
\cite{brands}. Most galaxies and 2D vortices are not described by
Tsallis polytropic distribution. A better model is a composite model
with an isothermal core where mixing is efficient ($q=1$) and a
polytropic halo ($q\neq 1$) where relaxation is incomplete \cite{grand}.

In conclusion, generalized entropies arise when {\it hidden
constraints} are in action: Casimir invariants, kinetic
constraints preventing mixing, forcing and dissipation, geometrical
structure of phase space (fractality) etc... We can either work with
the Boltzmann entropy and try to take into account these additional
constraints or keep only the usual constraints (mass and energy) and
change the form of entropy. The second possibility, while leading to
some indeterminations, is often more convenient.

\acknowledgements 

I thank F. Bouchet, J. Sommeria and C. Tsallis for stimulating
discussions and relevant comments. I also acknowledge interesting
discussions with O. Fliegans, R. Pasmanter, R. Robert, C. Rosier and
B. Turkington.

\appendix

\section{Linear stability analysis of the generalized Kramers equation}
\label{sec_lsK}

Let $f$ be a stationary solution of Eq. (\ref{k8}) and $\delta f$ a
small perturbation around this solution. Let us now linearize
Eq. (\ref{k8}) around equilibrium and write the time dependance of the
perturbation in the form $\delta f\sim e^{\lambda t}$. Noting that
$\nabla_{6}\cdot {\bf U}_{6}=0$, we get
\begin{equation}
\label{l1}
\lambda\delta f+\nabla_{6}(\delta f {\bf U}_{6})+\nabla_{6}(f \delta{\bf U}_{6})=-{\partial \delta {\bf J}_{f}\over\partial {\bf v}},
\end{equation}
where ${\bf J}_{f}$ is given by Eq. (\ref{k6}). Multiplying both side of Eq. (\ref{l1}) by $C''(f)\delta f$ and integrating over phase space, we obtain
\begin{eqnarray}
\label{l2}
\lambda\int C''(f)(\delta f)^{2}d^{3}{\bf r}d^{3}{\bf v}+\int C''(f)\delta f \nabla_{6}(\delta f{\bf U}_{6})d^{3}{\bf r}d^{3}{\bf v}+\nonumber\\
\int C''(f)\delta f\nabla_{6}(f\delta {\bf U}_{6})d^{3}{\bf r}d^{3}{\bf v}= -\int C''(f)\delta f {\partial \delta {\bf J}_{f}\over\partial {\bf v}}d^{3}{\bf r}d^{3}{\bf v}.
\end{eqnarray}
The second term can be rewritten
\begin{eqnarray}
\label{l3}
I_{2}={1\over 2}\int C''(f)\nabla_{6}\biggl\lbrack (\delta f)^{2}{\bf U}_{6}\biggr \rbrack d^{3}{\bf r}d^{3}{\bf v}=-{1\over 2}\int C'''(f) (\delta f)^{2}\nabla_{6} f\cdot {\bf U}_{6}d^{3}{\bf r}d^{3}{\bf v},
\end{eqnarray}
where we have used an integration by parts. From the relation (\ref{ge6}), we find
\begin{eqnarray}
\label{l4}
C''(f)\nabla_{6} f=-\beta {\bf U}_{6\perp},
\end{eqnarray}
Hence, $I_{2}=0$. Using $\delta {\bf U}_{6}=(0,-\nabla\delta\Phi)$ and Eq. (\ref{l4}), the third term in Eq. (\ref{l2}) can be rewritten
\begin{eqnarray}
\label{l5}
I_{3}=-\beta\int \delta f {\bf U}_{6\perp}\delta {\bf U}_{6} d^{3}{\bf r}d^{3}{\bf v}=\beta\int \delta f {\bf v}\cdot\nabla\delta \Phi d^{3}{\bf r}d^{3}{\bf v}.
\end{eqnarray}
After an integration by parts, the fourth term in Eq. (\ref{l2}) can be written
\begin{eqnarray}
\label{l6}
I_{4}=\int \delta {\bf J}_{f}\cdot {\partial\over\partial {\bf v}}\biggl\lbrack C''(f)\delta f\biggr\rbrack d^{3}{\bf r}d^{3}{\bf v},
\end{eqnarray}
or, using Eq. (\ref{l4}),
\begin{eqnarray}
\label{l7}
I_{4}=\int \delta {\bf J}_{f}\cdot \biggl\lbrack C''(f){\partial\delta f\over\partial {\bf v}}-\beta {C'''(f)\over C''(f)}\delta f{\bf v}\biggr\rbrack d^{3}{\bf r}d^{3}{\bf v}.
\end{eqnarray}
Taking the variation of Eq. (\ref{k6}) and using Eq. (\ref{l4}), one finds that
\begin{eqnarray}
\label{l8}
\delta{\bf J}_{f}=-Df \biggl\lbrack C''(f){\partial\delta f\over\partial {\bf v}}-\beta {C'''(f)\over C''(f)}\delta f{\bf v}+\delta\beta {\bf v}\biggr\rbrack.
\end{eqnarray}
Hence,
\begin{eqnarray}
\label{l9}
I_{4}=-\int \delta {\bf J}_{f}\cdot \biggl ({\delta{\bf J}_{f}\over Df}+\delta\beta {\bf v}\biggr)d^{3}{\bf r}d^{3}{\bf v}=-\int {(\delta {J}_{f})^{2}\over Df}d^{3}{\bf r}d^{3}{\bf v}.
\end{eqnarray}
where the last equality follows from the conservation of energy (\ref{k2}). 
Inserting the foregoing relations in Eq. (\ref{l2}), we obtain
\begin{eqnarray}
\label{l10}
\lambda\int C''(f)(\delta f)^{2}d^{3}{\bf r}d^{3}{\bf v}+\beta\int \delta f {\bf v}\cdot\nabla\delta \Phi d^{3}{\bf r}d^{3}{\bf v}=-\int {(\delta {J}_{f})^{2}\over Df}d^{3}{\bf r}d^{3}{\bf v}.
\end{eqnarray}
We now multiply both sides of Eq. (\ref{l1}) by $\delta\Phi$ and integrate over phase space. After straightforward integrations by parts, we get
\begin{eqnarray}
\label{l11}
\lambda\int \delta f\delta\Phi d^{3}{\bf r}d^{3}{\bf v}-\int \delta f {\bf v}\cdot\nabla\delta \Phi d^{3}{\bf r}d^{3}{\bf v}=0.
\end{eqnarray}
Combining this relation with Eq. (\ref{l10}), we find
\begin{eqnarray}
\label{l12}
\lambda\int C''(f)(\delta f)^{2}d^{3}{\bf r}d^{3}{\bf v}+\lambda\beta\int \delta f\delta\Phi d^{3}{\bf r}d^{3}{\bf v}=-\int {(\delta {J}_{f})^{2}\over Df}d^{3}{\bf r}d^{3}{\bf v}.
\end{eqnarray} 
Now, the left hand side is just proportional to the second order variations of the free energy $\delta^{2}{J}$, see Eq. (\ref{ts1}). On the other hand, recalling that ${\bf J}_{f}={\bf 0}$ at equilibrium, the second variations of the rate of entropy production (\ref{k11}) are given by
\begin{eqnarray}
\label{l13}
\delta^{2}\dot S=\int {(\delta J_{f})^{2}\over Df}d^{3}{\bf r}d^{3}{\bf v},
\end{eqnarray} 
and they are clearly positive. Therefore, Eq. (\ref{l12}) can be rewritten in the simple form (\ref{l15}).

\section{Linear stability analysis of the generalized Smoluchowski equation }
\label{sec_lsV}

Let $\omega$ be a stationary solution of
Eq. (\ref{g8}) and $\delta \omega$ a small perturbation around this
solution. Let us now linearize Eq. (\ref{g8}) around equilibrium and
write the time dependance of the perturbation in the form $\delta
\omega\sim e^{\lambda t}$. We get
\begin{equation}
\label{w1}
\lambda\delta \omega+\nabla (\delta \omega {\bf u})+\nabla (\omega \delta{\bf u})=-\nabla\delta {\bf J}_{\omega},
\end{equation}
where ${\bf J}_{\omega}$ is given by Eq. (\ref{g6}). Multiplying both side of Eq. (\ref{w1}) by $C''(\omega)\delta \omega$ and integrating over the whole domain, we obtain
\begin{eqnarray}
\label{w2}
\lambda\int C''(\omega)(\delta \omega)^{2}d^{2}{\bf r}+\int C''(\omega)\delta \omega \nabla (\delta \omega {\bf u})d^{2}{\bf r}+\nonumber\\
\int C''(\omega)\delta \omega\nabla (\omega\delta {\bf u})d^{2}{\bf r}= -\int C''(\omega)\delta \omega \nabla {\bf J}_{\omega} d^{2}{\bf r}.
\end{eqnarray}
The second term in Eq. (\ref{w2}) can be rewritten
\begin{eqnarray}
\label{w3}
I_{2}={1\over 2}\int C''(\omega)\nabla \biggl\lbrack (\delta \omega)^{2}{\bf u}\biggr \rbrack d^{2}{\bf r}=-{1\over 2}\int C'''(\omega) (\delta \omega)^{2}\nabla \omega\cdot {\bf u} d^{2}{\bf r},
\end{eqnarray}
where we have used an integration by parts. From the stationary condition (\ref{g12}), we obtain 
\begin{eqnarray}
\label{w4}
C''(\omega)\nabla \omega=-\beta \nabla\psi.
\end{eqnarray}
Since ${\bf u}\cdot\nabla\psi=0$ we conclude that $I_{2}=0$.  Using Eq. (\ref{w4}), the third term in Eq. (\ref{w2}) can be rewritten
\begin{eqnarray}
\label{w5}
I_{3}=-\beta\int \delta \omega \delta {\bf u}\nabla\psi d^{2}{\bf r}=\beta\int \delta \omega {\bf u}\nabla\delta\psi d^{2}{\bf r}.
\end{eqnarray}
After an integration by parts, the fourth term can be written
\begin{eqnarray}
\label{w6}
I_{4}=\int \delta {\bf J}_{\omega}\cdot \nabla \biggl\lbrack C''(\omega)\delta \omega\biggr\rbrack d^{2}{\bf r},
\end{eqnarray}
or, using Eq. (\ref{w4}),
\begin{eqnarray}
\label{w7}
I_{4}=\int \delta {\bf J}_{\omega}\cdot \biggl\lbrack C''(\omega)\nabla\delta\omega-\beta {C'''(\omega)\over C''(\omega)}\delta \omega \nabla\psi\biggr\rbrack d^{2}{\bf r}.
\end{eqnarray}
Taking the variation of Eq. (\ref{g6}) and using Eq. (\ref{w4}), we find that
\begin{eqnarray}
\label{w8}
\delta{\bf J}_{\omega}=-D\omega \biggl\lbrack C''(\omega)\nabla\delta\omega-\beta {C'''(\omega)\over C''(\omega)}\delta \omega \nabla\psi+\delta\beta \nabla\psi+\beta\nabla\delta\psi\biggr\rbrack.
\end{eqnarray}
Hence,
\begin{eqnarray}
\label{w9}
I_{4}=-\int \delta {\bf J}_{\omega}\cdot \biggl ({\delta{\bf J}_{\omega}\over D\omega}+\beta \nabla\delta\psi+\delta\beta\nabla\psi \biggr)d^{2}{\bf r}=-\int \biggl ({\delta{J}_{\omega}^{2}\over D\omega}+\delta {\bf J}_{\omega}\beta \nabla\delta\psi\biggr)d^{2}{\bf r},
\end{eqnarray}
where we have used the conservation of energy (\ref{g3}) to get the last equality. Inserting these results in Eq. (\ref{w2}), we obtain
\begin{eqnarray}
\label{w10}
\lambda\int C''(\omega)(\delta \omega)^{2}d^{2}{\bf r}+\beta\int \delta \omega {\bf u}\nabla\delta \psi d^{2}{\bf r}=-\int \biggl ({\delta{J}_{\omega}^{2}\over D\omega}+\delta {\bf J}_{\omega}\beta \nabla\delta\psi\biggr)d^{2}{\bf r}.
\end{eqnarray}
We now multiply both sides of Eq. (\ref{w1}) by $\delta\psi$ and integrate over the domain. After straightforward integrations by parts, and using $\delta{\bf u}\cdot\nabla\delta\psi=0$, we find
\begin{eqnarray}
\label{w11}
\lambda\int \delta \omega\delta\psi d^{2}{\bf r}-\int \delta\omega {\bf u}\cdot \nabla\delta\psi d^{2}{\bf r}=\int \delta {\bf J}_{\omega}\cdot \nabla\delta\psi d^{2}{\bf r}.
\end{eqnarray}
Combining with Eq. (\ref{w10}), we get
\begin{eqnarray}
\label{w12}
\lambda\int C''(\omega)(\delta \omega)^{2}d^{2}{\bf r}+\lambda\beta\int \delta \omega\delta\psi d^{2}{\bf r}=-\int {(\delta {J}_{\omega})^{2}\over D\omega}d^{2}{\bf r}.
\end{eqnarray} 
Now, the left hand side is just proportional to the second order variations of the free energy $\delta^{2}{J}$, see Eq. (\ref{v10}). On the other hand, recalling that ${\bf J}_{\omega}={\bf 0}$ at equilibrium, the second order variations of the rate of entropy production (\ref{g10}) are given by
\begin{eqnarray}
\label{w13}
\delta^{2}\dot S=\int {(\delta J_{\omega})^{2}\over D\omega}d^{2}{\bf r},
\end{eqnarray} 
and they are clearly positive (see discussion after
Eq. (\ref{g10})). Therefore, Eq. (\ref{w12}) can be rewritten in the
simple form (\ref{l15}).

\section{Sufficient conditions of stability and Arnold theorems}
\label{sec_suff}

According to Eqs. (\ref{new}) and (\ref{v10}), $\delta^{2}J$ can be
rewritten
\begin{eqnarray}
\label{v12}
\delta^{2}J=-{\beta\over 2}\biggl\lbrace \int {(\delta \omega)^{2}\over -\omega'(\psi)}d^{2}{\bf  r}+\int \delta\omega\delta\psi d^{2}{\bf  r}\biggr\rbrace.
\end{eqnarray}
The term is bracket is called Arnold pseudo-energy or Arnold
invariant. Using an integration by parts, we have equivalently
\begin{eqnarray}
\label{suff1}
\delta^{2}J=-{\beta\over 2}\biggl\lbrace \int {(\delta \omega)^{2}\over -\omega'(\psi)}d^{2}{\bf  r}+\int (\nabla\delta\psi)^{2} d^{2}{\bf  r}\biggr\rbrace.
\end{eqnarray}
First assume that $\beta>0$. By Eq. (\ref{new}), we see that
$\omega'(\psi)<0$. Therefore, $\delta^{2}J<0$ and the system is
stable. If $\beta>0$, then $\omega'(\psi)>0$ and we cannot conclude
directly. Let us introduce a set of normalized eigenfunctions
$\phi_{i}({\bf r})$ such that $-\Delta \phi_{i}=\lambda_{i}\phi_{i}$
with $\phi_{i}=0$ on the boundary of the domain and $\int
\phi_{i}\phi_{j}d^{2}{\bf r}=\delta_{ij}$. Noting that $-\int
\phi_{i}\Delta\phi_{i}d^{2}{\bf r}=\lambda_{i}$ and integrating by
parts, we find that $\lambda_{i}=\int (\nabla\phi_{i})^{2}d^{2}{\bf
r}>0$. Then, we decompose $\delta\omega$ and $\delta\phi$ on these
eignefunctions. Using the Poisson equation $\omega=-\Delta\psi$, we have
$\delta\omega=\sum_{i}\delta\omega_{i}\phi_{i}$ and
$\delta\psi=\sum_{i}{\delta\omega_{i}\over\lambda_{i}}\phi_{i}$. Therefore,
\begin{eqnarray}
\label{suff2}
\int \delta\omega\delta\psi d^{2}{\bf r}=\sum_{i}{(\delta\omega_{i})^{2}\over \lambda_{i}}.
\end{eqnarray}
We now label the eigenvalues such that $0<\lambda_{1}<\lambda_{2}<...$. Then 
$\lambda_{i}\ge \lambda_{1}$ and we get  
\begin{eqnarray}
\label{suff3}
\int \delta\omega\delta\psi d^{2}{\bf r}\le {1\over\lambda_{1}}\sum_{i}{(\delta\omega_{i})^{2}}={1\over\lambda_{1}}\int (\delta\omega)^{2} d^{2}{\bf r}.
\end{eqnarray}
Therefore,
\begin{eqnarray}
\label{suff4}
\delta^{2}J\le -{\beta\over 2}\int \biggl \lbrack {1\over\lambda_{1}}-{1\over \omega'(\psi)}\biggr \rbrack (\delta\omega)^{2} d^{2}{\bf  r}.
\end{eqnarray}
As a result, if $0\le \omega'(\psi)<\lambda_{1}$, then $\delta^{2}J\le
0$ and the system is stable. Using Eq. (\ref{new}), this condition can
be rewritten $\beta>-\lambda_{1}{\rm min}\ C''(\omega)$. These
sufficient conditions of stability are called Arnold theorems. They
imply canonical and microcanonical stability in the thermodynamical
analogy.  Our general formalism shows that the Arnold theorems can
have applications in many other situations than just fluid dynamics
(see, e.g., \cite{grand}).

\end{document}